\documentclass[lettersize,journal]{IEEEtran}
\usepackage{amsmath,amsfonts}
\usepackage{algorithmic}
\usepackage{algorithm}
\usepackage{array}
\usepackage[caption=false,font=normalsize,labelfont=sf,textfont=sf]{subfig}
\usepackage{textcomp}
\usepackage{stfloats}
\usepackage{url}
\usepackage{verbatim}
\usepackage{graphicx}
\usepackage{cite}
\usepackage[colorlinks=false, pdfborder={0 0 0}]{hyperref}  

\begin{document}

\title{Learning to Compress and Transmit: Adaptive Rate Control for Semantic Communications over LEO Satellite-to-Ground Links}

\author{Jiangtao~Luo,~\IEEEmembership{Senior Member,~IEEE}, Yongyi~Ran,~\IEEEmembership{Member,~IEEE}, Guoliang Xu and Jihua Zhou
\thanks{Jiangtao Luo, Yongyi Ran and Guoliang Xu are with the School of Communications and Information Engineering, Chongqing University of Posts and Telecommunications, Chongqing, 400065 China. Jihua  Zhou is with the College Of Computer and Information Science, Southwest University, Chongqing 400715, China. Jiangtao Luo and Yongyi Ran are the corresponding authors. (Email: Luojt@cqupt.edu.cn, ranyy@cqupt.edu.cn). 
}
\thanks{This work is supported by the National Natural Science Foundation of China (No.~U25B2033, No.~62171072 and No.~U23A20275).}
}



\maketitle

\begin{abstract}
The bottleneck of satellite-to-ground links poses a major challenge for the timely downlink of massive on-board imagery. This paper studies adaptive image transmission over LEO satellite-to-ground links using joint source–channel coding (JSCC). We propose an RL-based framework that dynamically selects the channel dimension (compression ratio) of a SwinJSCC encoder to maximize the number of receivedes satisfying reconstruction-quality constraints (PSNR and MS-SSIM) within a finite visibility window. The agent leverages SNR prediction to perform proactive rate adaptation and incorporates an on-board transmission-queue model that captures bursty encoding while penalizing both buffer overflow and underutilization. Simulations under realistic overpass conditions show that the proposed policy substantially outperforms fixed-rate baselines, achieving nearly 95\% qualified frames with zero packet loss.
\end{abstract}


\section{Introduction}
\label{sec:intro}

  \IEEEPARstart{T}{he} rapid growth of data generated aboard low Earth orbit (LEO) satellites poses a critical challenge: how to efficiently transmit this data back to the ground. Traditional pixel-wise compression methods, such as JPEG2000, BPG, and emerging deep learning-based approaches including DeepSpace~\cite{SunCH25-DeepSpace-SigComm}, have approached their fundamental limits yet still struggle to meet the increasing demands of high-throughput satellite missions. On the other hand, on-board artificial intelligence (AI) processing offers an alternative, but remains severely constrained by the extremely limited computational and energy resources available on satellite platforms.

Semantic communication~\cite{ZhangPing22-TowardsWisdom-Engineering} has emerged as a promising paradigm to address this challenge. Instead of transmitting raw pixels, a semantic encoder extracts and transmits compact semantic features to the ground, where a decoder reconstructs the original image. This approach offers two key advantages. First, it significantly reduces the amount of data to be transmitted, alleviating the bottleneck of limited downlink bandwidth. Second, semantic communication exhibits inherent robustness to channel variations, 
effectively overcoming the well-known ``cliff effect'' that plagues conventional 
digital communication systems. This makes it particularly suitable for dynamic LEO satellite-to-ground links, where channel conditions vary rapidly over time.

A prominent implementation of semantic communication is deep joint source-channel coding (DeepJSCC)~\cite{Bourtsoulatze19-DeepJSCC-TCCN}, which integrates source coding, channel coding, and modulation into a single end-to-end optimized neural network. Unlike traditional separation-based schemes that optimize source and channel modules independently, DeepJSCC learns a joint representation that is directly mapped to channel input symbols. This approach not only eliminates the cliff effect by gracefully degrading reconstruction quality as channel conditions worsen, but also achieves superior performance at low signal-to-noise ratios (SNR) and limited bandwidth regimes. Building upon this foundation, \cite{YangKe25-SwinJSCC-TCCN} proposed SwinJSCC, which replaces the convolutional neural network (CNN) backbone with a Swin Transformer architecture. The SwinJSCC encoder leverages hierarchical feature extraction and window-based self-attention mechanisms to capture long-range dependencies in images, leading to improved reconstruction quality, especially for high-resolution imagery. Furthermore, SwinJSCC incorporates two plug-in modules -- \textit{Channel} ModNet and \textit{Rate}  ModNet -- that enable a single model to adapt flexibly to varying channel SNR and transmission rates, making it particularly attractive for dynamic satellite communication scenarios.

Despite its potential, research on semantic communication for satellite links, especially for LEO feeder links, remains limited. Several critical aspects unique to satellite communications have been largely overlooked in existing work. First, most semantic communication frameworks assume an idealized setting where the encoder has instantaneous and perfect knowledge of the channel state, and where the transmission itself incurs negligible delay. In practice, however, LEO satellite links exhibit non-negligible propagation delays that vary with the elevation angle, typically ranging from several to tens of milliseconds. These delays, combined with the uplink command transmission and on-board processing, introduce a control loop latency that can render decisions based on current channel observations outdated by the time the corresponding signal is actually transmitted. Second, existing studies rarely consider the impact of on-board transmission buffering. In a satellite system, encoded symbols must be queued before transmission, and the queue dynamics -- influenced by both the selected compression ratio and the time-varying channel symbol rate -- directly affect the system's ability to avoid buffer overflow and maximize throughput. To the best of our knowledge, no prior work on semantic communication has systematically addressed the coupled challenges of propagation-induced control latency and queue-aware rate adaptation for LEO satellite links. 

In this paper, we target the efficient downlink transmission of observed images from LEO satellites to ground gateways. We propose an adaptive semantic communication framework based on SwinJSCC, a joint source-channel coding scheme built upon the Swin Transformer architecture. The objective is to maximize the number of successfully transmitted images that meet predefined quality requirements (in terms of PSNR and MS-SSIM) within a single satellite overpass. To cope with the time-varying channel conditions and control loop latency, we further integrate a reinforcement learning agent with SNR prediction to dynamically adjust the compression ratio of the SwinJSCC encoder. The proposed framework operates in a \emph{predict–decide–execute} paradigm, where the ground gateway forecasts future channel states, selects an optimal compression ratio, and sends the decision to the satellite for execution. Extensive simulations demonstrate that our approach significantly outperforms fixed-rate and threshold-based strategies in terms of both throughput and image quality.

The main contributions of this paper are summarized as follows:
\begin{itemize}
    \item \textbf{Queue-aware and latency-resilient rate control:} We identify two critical but overlooked challenges in semantic communication for LEO satellites: elevation-dependent propagation delays that induce control loop latency, and on-board transmission queue dynamics that affect throughput. To address these, we introduce a \emph{predict–decide–execute} framework that integrates SNR predictors for future SNR leveraging link budget, enabling the RL agent to select compression ratios aligned with future channel and buffer states.
    
    \item \textbf{Ground-based deployment with closed-loop telemetry:} We propose a practical deployment architecture where the RL agent and SNR predictor reside at the ground gateway. The satellite periodically downlinks queue length telemetry, while the gateway estimates SNR from received signals. Uplink commands deliver both the selected compression ratio and the predicted SNR to the satellite, where the Channel ModNet and Rate ModNet update accordingly.
    
    \item \textbf{Comprehensive simulation and performance evaluation:} We build a realistic LEO feeder link simulator incorporating elevation-dependent SNR trajectories and variable propagation delays. Extensive overpass-level simulations demonstrate that our approach increases the number of successfully transmitted images by up to 5\% compared to fixed-rate baselines strategies, while maintaining stable queue occupancy and meeting quality constraints.
\end{itemize}

\section{Related work}
\label{sec:related}
\subsection{Joint Source-Channel Coding for Image Transmission}

Deep joint source-channel coding (DeepJSCC) has emerged as a promising paradigm for wireless image transmission, eliminating the ``cliff effect'' inherent in separate source and channel coding. \cite{Bourtsoulatze19-DeepJSCC-TCCN} established the theoretical foundation for semantic communication networks, paving the way for practical DeepJSCC implementations. Building on this foundation, \cite{XuJL22-WiressAttension-ADJSCC-TCSVT} proposed ADJSCC which introduced attention modules into DeepJSCC, enabling the model to adaptively focus on informative features and significantly improving its performance, particularly at low signal-to-noise ratios (SNR) where it outperforms conventional separation-based schemes. Subsequently, \cite{YangKe25-SwinJSCC-TCCN} proposed SwinJSCC, which replaces the CNN backbone with a Swin Transformer, achieving superior performance for high-resolution images through hierarchical feature extraction and window-based self-attention. SwinJSCC also introduces Channel ModNet and Rate ModNet modules, enabling a single model to adapt to varying channel conditions and transmission rates.

\subsection{Related Work on Adaptive Deep JSCC for Satellite Communications}

Existing adaptive deep joint source-channel coding (JSCC) methods for satellite image transmission can be broadly categorized into two families based on their backbone architecture: convolutional neural network (CNN) based and Vision Transformer (ViT) based. Each family shares common limitations that are particularly critical for LEO satellite links.

\paragraph{CNN-based Approaches.}
Several works adopt CNN-based DeepJSCC as their foundation. \cite{GuoYL24-SatelliteTransmitWithJSCC-WCSP} formulates a resource allocation problem for DeepJSCC-based satellite transmission, jointly optimizing image resolution, information entropy, latency, and power via linear programming. \cite{JiangYan25-HighQualityCompression-TSC} proposes ASE-JSCC, which jointly optimizes remote sensing image transmission with downstream classification tasks, achieving high compression ratios (up to 384$\times$) through semantic feature selection and noise-injected training. Most recently, \cite{TanZYY26-AdaptiveSCforRemoteSensing-TWC} presents ARJSCC, which adapts semantic feature extraction to varying SNR via an attention module, adjusts semantic length via a position mask, and transmits a compressed residual with a dedicated enhancement model.

Despite their innovations, all CNN-based approaches share a fundamental limitation: the inherently limited receptive field and model capacity of CNNs, which struggle to capture long-range dependencies in high-resolution remote sensing images. Moreover, they commonly assume instantaneous channel feedback and negligible transmission delay, overlooking the elevation-dependent propagation delays and the timeliness requirements of real-time data delivery. On-board queue dynamics, which directly affect throughput and delay in satellite systems, are also ignored. Consequently, these methods are ill-suited for dynamic LEO satellite links where control latency and buffer management are critical.

\paragraph{ViT-based Approaches.}
To overcome the capacity constraints of CNNs, some works turn to Vision Transformers. \cite{LiZY25-ChannelAdaptiveSemSatCommRS-ICC} proposes SatSemCom, which features a ViT-based semantic encoder, a channel predictor to mitigate outdated CSI, and a rate adaptation module. \cite{YinYB25-JSCC-MultiModalGSL-WCNC} introduces a ViT-based JSCC approach for multi-modal satellite-to-ground communication, learning common semantic information across modalities to reduce coding space.

However, ViT-based encoders suffer from quadratic computational complexity with respect to image resolution, limiting their scalability to high-resolution imagery (e.g., the reported input size in \cite{YinYB25-JSCC-MultiModalGSL-WCNC} is only $3 \times 512 \times 512$). In contrast, hierarchical architectures like Swin Transformer achieve linear complexity via shifted window attention. Furthermore, like their CNN counterparts, these ViT-based methods also neglect on-board queue dynamics and elevation-dependent propagation delays, which are essential for real-time adaptive transmission over LEO links.

\paragraph{Predictive and Adaptive Coding.}
\cite{ZhangWY23-PredictiveAdaptiveDeepCoding-TWC} proposes PADC, a predictive framework that minimizes transmission rate under PSNR constraints by predicting reconstruction quality from image content, SNR, and compression ratio. While effective in general settings, its transmitter-side prediction module imposes on-board computation that is challenging for resource-constrained LEO satellites. It also assumes instantaneous feedback and does not address control latency or queue dynamics.

In summary, existing CNN-based JSCC methods lack the capacity to handle high-resolution imagery, while ViT-based methods face quadratic complexity. Nearly all assume instantaneous feedback and ignore propagation-induced latency, on-board queue dynamics, and adaptive rate control over a complete satellite overpass. These gaps motivate our work, which combines a Swin Transformer-based JSCC encoder (offering linear complexity) with a ground-based RL agent, link-budget based SNR prediction, and a queue-aware reward design tailored for LEO satellite links.

\subsection{Positioning of This Work}

In contrast to the above, existing semantic and joint source-channel coding methods for satellite image transmission suffer from several common limitations: limited model capacity (CNN-based), quadratic complexity (ViT-based), neglect of propagation delays and on-board queue dynamics, transmitter-side predictive burden, and lack of joint optimization over a full overpass.

This paper advances the state of the art by (i) adopting SwinJSCC with linear complexity for high-resolution multi-spectral images, (ii) explicitly addressing control loop latency via SNR prediction in a predict–decide–execute paradigm, (iii) incorporating queue dynamics into RL state and reward for proactive congestion control, (iv) deploying the RL agent and predictor on the ground to minimize on-board computation, and (v) optimizing over a complete satellite overpass to maximize qualified frames under quality constraints. Simulations demonstrate significant gains over fixed-rate baselines.

\section{System Model and Problem Formulation}
\label{sec:model}
%

\subsection{System Model}
\begin{figure}
    \centering
    \includegraphics[width=0.95\linewidth]{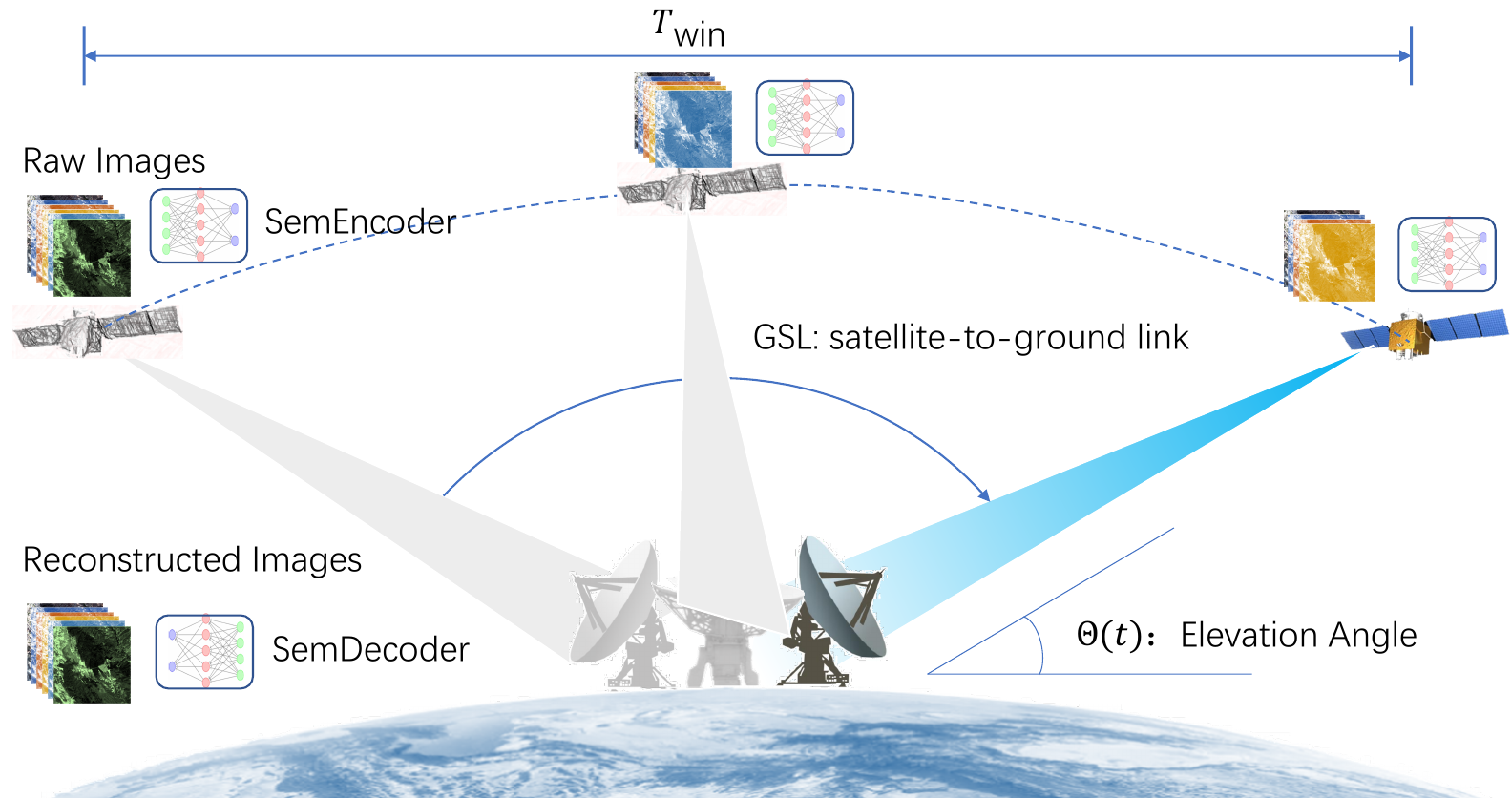}
    \caption{System architecture of semantic communnications for multispectral images over GSL.}
    \label{fig:sys_model}
\end{figure}
The system model is illustrated in Fig.~\ref{fig:sys_model}. We consider a low Earth orbit (LEO) Earth observation satellite transmitting captured images to a ground gateway. To cope with the dynamic loss characteristics of the satellite-to-ground link (GSL), a semantic encoder (SemEncoder) based on the SwinJSCC scheme is deployed on board. Specifically, each original image is first semantically encoded by the SemEncoder, and the resulting symbols are placed into a transmission queue before being sent over the GSL. Upon reception, the ground gateway reconstructs the image using the corresponding semantic decoder (SemDecoder).


The GSL considered in this work is the feeder link from the satellite to the ground gateway. We adopt the additive white Gaussian noise (AWGN) channel as the baseline model, augmented with a constant additional attenuation margin (e.g., $2.5$ dB) to account for simplified atmospheric effects, rather than explicitly modeling rain fading or tropospheric scintillation. More realistic random or bursty impairments can be incorporated in future extensions. The variation of SNR with satellite elevation angle over a single overpass is explicitly characterized. Unlike the original SwinJSCC studies that assume an idealized channel with stationary statistics, our work introduces two key practical considerations: an on-board transmission queue to model buffer dynamics, and the time-varying SNR trajectory over the satellite visibility window, which captures the realistic evolution of link quality during an overpass.

\subsection{Channel Model for LEO Feeder Link}

The LEO satellite-to-gateway (feeder) downlink is modeled as an additive white Gaussian noise (AWGN) channel with time-varying signal-to-noise ratio (SNR). The received power $P_r(t)$ varies deterministically with the satellite elevation angle $\theta(t)$ over a visibility window of approximately $600$ seconds (assuming a $900$ km orbit altitude), following:
\[
P_r(t) = \underbrace{(P_t + G_t)}_{\text{EIRP}} + G_r - L, \quad \text{(dB)},
\]
where $P_t$ is the satellite transmit power (dBW), $G_t$ and $G_r$ are the antenna gains of the satellite and the ground gateway (dBi), respectively, and $L = L_{\text{FSPL}}(t) + L_{\text{extra}}(t)$ denotes the total loss (dB), which includes the free-space path loss $L_{\text{FSPL}}(t)$ and all additional atmospheric losses $L_{\text{extra}}(t)$ (such as gas absorption, rain attenuation, and other impairments). 
The instantaneous SNR, denoted as $\gamma(t)$, is given by $\gamma(t) = P_r(t) - N_0 - 10\log_{10}(B_n)$, where $N_0$ is the noise power spectral density (in dBW/Hz or dBm/Hz), and $B_n$ is the receiver noise bandwidth (in Hz). 

The AWGN simplification is justified by the gateway’s high-gain directional antenna and open-area deployment, which effectively eliminate multipath propagation. In an AWGN channel, the multiplicative distortion is a purely amplitude-scaling factor $\sqrt{P_r(t)}$, which is \emph{deterministic and known} from the satellite's trajectory. This is fundamentally different from a fading channel, where the amplitude scaling is a \emph{random, unknown} process. The received complex baseband signal is therefore expressed as:
\[
y(t) = \sqrt{P_r(t)} \cdot x(t) + n(t), \quad n(t) \sim \mathcal{CN}(0, N_0).
\]

In engineering practice, the link performance of the satellite-to-ground feeder link is preferred to be compactly expressed using the satellite effective isotropic radiated power (EIRP), the ground station figure of merit $G/T$, and the carrier-to-noise density ratio $C/N_0$. The fundamental link equation for $C/N_0$ is given by:
\begin{equation}\label{eqn:CN0}
C/N_0 = \text{EIRP} - L + G/T - 10\log_{10}(k) \quad \text{(dBHz)},
\end{equation}
where $k = 1.380649 \times 10^{-23}$ J/K is Boltzmann's constant, and $10\log_{10}(k) \approx -228.6$ dBW/K/Hz. 
The instantaneous signal-to-noise ratio $\gamma(t)$ is then obtained from $C/N_0$ by accounting for the receiver noise bandwidth $B_n$:
\begin{equation}\label{eqn:snr}
\gamma(t) = C/N_0 - 10\log_{10}(B_n) \quad \text{(dB)}.
\end{equation}
In this paper, this formulation directly links the engineering parameters (EIRP, $G/T$) to the SNR used in our channel model, providing a practical and modular approach for link budget calculations and ensuring alignment with realistic engineering parameters.


\subsection{Symbol-Level Transmit Queue Model}
\begin{figure}
    \centering
    \includegraphics[width=0.95\linewidth]{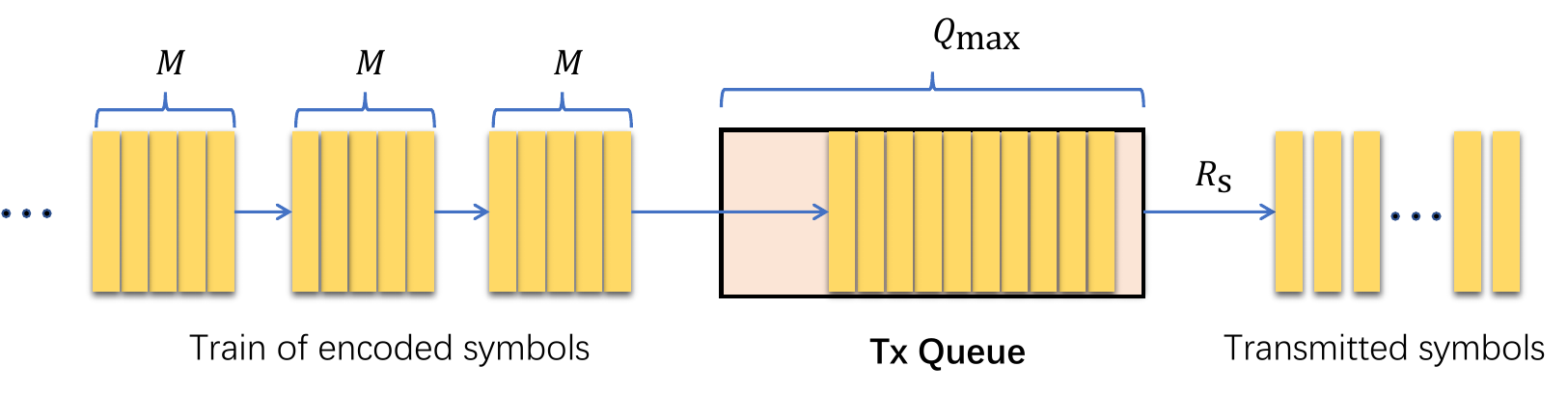}
    \caption{Transmit queue model.}
    \label{fig:tx_queue}
\end{figure}
To bridge the gap between the bursty output of the semantic encoder and the constant-rate physical layer, we introduce a transmission queue for encoded symbols at the satellite. This queue absorbs short-term fluctuations in symbol generation and provides a realistic abstraction of on-board buffering constraints.

\subsubsection{Queue Dynamics} 
To improve computational efficiency and fully exploit the parallel processing capabilities of neural network accelerators (e.g., GPUs or on-board NPUs), we adopt a batch-wise encoding strategy that naturally matches the bursty nature of semantic encoding. Processing a batch of images in a single forward pass significantly reduces the number of kernel launches and memory transfers compared to sequential single-image processing, thereby lowering inference latency and on-board energy consumption. Let $T_d$ denote the decision interval, which defines the period over which the agent selects a compression ratio. Within each decision interval, the encoder processes a batch of $M$ images, as illustrated in Fig.~\ref{fig:tx_queue}. For a given compression ratio $R_t$, the number of symbols generated for a single image is $N_{\text{sym}}(R_t)$; consequently, a batch of $M$ images produces a total of $A_t = M \times N_{\text{sym}}(R_t)$ symbols. After encoding, these symbols are pushed into the transmission queue sequentially, one image at a time.

The physical layer transmits symbols at a constant rate $R_s$ (symbols per second), which is determined by the signal bandwidth $B_s$ and the pulse shaping filter. Specifically, for a system with bandwidth $B_s$ and a raised-cosine filter with roll-off factor $\alpha$, the symbol rate is given by $R_s = B_s / (1 + \alpha)$. Under ideal low-pass filtering ($\alpha \to 0$) and no channel distortions, we have the approximation $R_s \approx B_s$.

Let $Q_t$ denote the queue length (in symbols) at the beginning of time slot $t$. The queue evolves as symbols are added from encoded images and removed by the physical layer at rate $R_s$. The number of symbols that can be transmitted during one decision interval of length $T_d$ is $R_s T_d$. Thus, the queue update is given by:
\[
Q_{t+1} = \max\left(0,\; Q_t + A_t - R_s \cdot T_d\right),
\]
where $\max(\cdot)$ ensures that the queue length never becomes negative (idle periods are allowed), and $A_t$ is the total number of symbols generated in time slot $t$.

\subsubsection{Two Queue Indicators}

The queue update equation $Q_{t+1} $ presented earlier (assuming an infinite buffer) provides an idealized evolution assuming an infinite buffer. In practice, the satellite has finite on-board memory, which imposes a hard upper bound on the queue length. This maximum queue capacity, denoted $Q_{\max}$, must be explicitly defined to capture buffer overflow and the resulting image discards.

We introduce two indicators to characterize the transmission queue behavior:

\begin{itemize}
    \item \textbf{Queue Capacity Indicator (QCI), $q_c$}: 
    The maximum queue length $Q_{\max}$ is set to accommodate the maximum symbol arrival over $q_c$ consecutive decision intervals under the highest compression ratio:
    \[
    Q_{\max} = q_c \times M \times N_{\text{sym}}^{\max},
    \]
    where $N_{\text{sym}}^{\max}$ is the number of symbols per image when the maximum allowed compression ratio (or channel number) is selected. The parameter $q_c$ controls the buffer depth in units of decision intervals, balancing robustness against bursty arrivals against on-board memory constraints.

    \item \textbf{Queue Drain Indicator (QDI), $q_d$}: 
    This indicator represents the number of decision intervals required to completely empty a full queue. The drain rate is determined by the physical layer symbol rate $R_s$ and the decision interval length $T_d$,  hence:
    \[
    q_d = \left\lceil \frac{Q_{\max}}{R_s T_d} \right\rceil.
    \]
    $q_d$ reflects the time needed to recover from a worst-case buffer occupancy and measures the system's responsiveness under congestion.
\end{itemize}

Specifically, using the Kodak image dimensions ($768 \times 512$ pixels, RGB) and a maximum channel number $C_K = 192$, we compute $N_{\text{sym}}^{\max} = 147,\!456$ symbols per image. With $q_c = 3$ and $M = 15$, the queue capacity is $Q_{\max} = 6,\!635,\!520$ symbols. Setting $q_d = 6$ yields a per‑decision‑interval transmission budget of $S_{\text{tx}} = R_s \cdot T_d \approx 1.11 \times 10^6$ symbols. For a decision interval $T_d = 5$ s, the required physical layer symbol rate is $R_s \approx 0.22$ Msym/s; for $T_d = 1$ s, $R_s \approx 1.11$ Msym/s. These values confirm the feasibility of the proposed queue model under realistic LEO satellite link budgets.

\subsubsection{Discard Policy}
When pushing the $M$ encoded images of a batch into the queue, we check the available space before adding each image. For the $j$-th image in the batch ($j = 1, \ldots, M$), if the current queue length $Q$ satisfies $Q + N_{\text{sym}}(R_t) \le Q_{\max}$, the image is successfully queued. Otherwise, the image is discarded, and no further images from the same batch are queued (i.e., discarding is sequential and stops at the first failure). Consequently, a batch may result in $d$ discarded images, where $0 \le d \le M$, and only $M - d$ images are successfully transmitted in that decision interval. This per-image discard policy preserves the semantic integrity of each transmitted image — an image is either fully queued or completely discarded — while allowing partial batch transmission to maximize throughput under congestion.

To discourage discards, the RL agent receives a penalty proportional to the number of discarded images $d$ in each decision interval. This incentivizes the agent to proactively reduce the compression ratio $R_t$ when the queue length approaches $Q_{\max}$, thereby avoiding buffer overflow and ensuring reliable transmission.

For long-term stability, the average arrival rate must not exceed the average service rate:
\[
\frac{1}{2}M \times P \times \mathbb{E}[R_t] < R_s \times T_d,
\]
where $\mathbb{E}[R_t]$ is the expected compression ratio under the learned policy. This condition is naturally enforced by the RL reward design, which penalizes queue buildup and discards.

The queue serves three critical functions in our framework. First, it decouples the semantic encoder (which operates in bursts) from the physical layer (which transmits continuously), allowing each to operate at its own pace. Second, the queue length $Q_t$ provides a direct measure of congestion, which is fed back to the RL agent via the telemetry downlink. By observing $Q_t$, the agent learns to reduce the compression ratio $R_t$ when the queue builds up, and to increase it when the queue is empty, thereby balancing transmission efficiency and buffer occupancy. Third, the per-image discard policy preserves semantic integrity, while the associated penalty guides the agent toward congestion-aware rate adaptation.

\subsection{Problem Formulation}
We consider a LEO satellite downlink communications system where a satellite transmits a sequence of images to a ground gateway during a single visiblility window of duration $T_{\text{win}}$ seconds. The satellite is equipped with a SwinJSCC encoder that compresses each image into a variable-length complex symbol vector based on a comression ratio $R \in \mathcal{R}$, where $\mathcal{R} = \{R_1, R_2, \ldots, R_K\}$ is a discrete set of admissible ratios.

The communication channel is time-varying due to satellite motion, atmospheric effects (rain fading, scintillation), and potential interference. At each time step $t$, the channel quality is characterized by the signal-to-noise ratio $\text{SNR}(t)$, which depends on the satellite elevation angle $\theta(t)$ and stochastic environmental conditions.

\subsubsection{System Dynamics}
Let $\mathcal{T} = \{1, 2, \ldots, T\}$ denote the set of decision epochs, where $T$ is the total number of decision steps within the visibility window (e.g., each step corresponds to one second). At each epoch $t$, the system is described by the following state vector:
\begin{equation}\label{eqn:state}
    \mathbf{s}_t = (\gamma_t,\ \theta_t,\ Q_t,\ R_{t-1}) \in \mathcal{S},
\end{equation}
where $\gamma_t \in \mathbb{R}^+$ is the estimated SNR at time step $t$; $\theta_t \in [\theta_{\text{min}}, 90^{\circ}]$ is the satellite elevation angle; $Q_t \in [0, Q_{\text{max}}]$ is the current length of the transmission buffer (in symbols); $R_{t-1} \in \mathcal{R}$ is the compression ratio selected at the previous decision epoch.


\subsubsection{Action Space}
The agent selects a discrete action at each decision epoch. Let $\mathcal{R} = \{R_1, R_2, \ldots, R_K\}$ denote the finite set of admissible compression ratios, where $K$ is the number of discrete choices. For each $k \in \{1, \ldots, K\}$, $R_k \in (0, 1]$ represents a specific compression ratio supported by the SwinJSCC encoder. 

At decision epoch $t$, the agent selects an action $a_t = R_t \in \mathcal{R}$, where $R_t$ is the target bandwidth ratio (compression ratio) to be applied to the next image. The action space is therefore $\mathcal{A} = \mathcal{R}$, with cardinality $|\mathcal{A}| = K$.

When action $a_t = R_k$ is selected, the number of complex symbols generated by the SwinJSCC encoder for the next image is determined by the bandwidth ratio $R_k$ and the total number of pixels in the original image. Let $P_{\text{in}} = H \times W \times C_{\text{in}}$ denote the total number of input pixels per image, where $C_{\text{in}} = 3$ for RGB images. The number of transmitted symbols per encoded image is then given by:
\begin{equation}
\label{eq:symbol_counts}
    N_{\text{sym}}(R_t) = \frac{P_{\text{in}}}{2} \times R_t = \frac{1}{2} \times \frac{H}{2^i} \times \frac{W}{2^i} \times C_t = S_{\text{iHW}} \times C_t,
\end{equation}
where $i$ is the number of stages in the SwinJSCC encoder, and $C_t \in \mathcal{C} = \{C_1, C_2, \ldots, C_K\}$ denotes the number of channels selected by the agent for the Rate ModNet at decision step $t$, i.e., $C_t$ corresponds to the compression ratio $R_t$. Here, $S_{\text{iHW}} = \frac{1}{2} \times \frac{H}{2^i} \times \frac{W}{2^i}$ is a constant determined by the input image dimensions and the number of stages in the SwinJSCC encoder. Specifically, following the 4-stage SwinJSCC encoder~\cite{YangKe25-SwinJSCC-TCCN}, there are five levels of channel numbers, i.e., $\mathcal{C} = \{32, 64, 96, 128, 192\}$, which correspond to the compression ratio set $\mathcal{R} = \{1/48, 1/24, 1/16, 1/12, 1/8\}$. This formulation directly follows the definition of the bandwidth ratio in the original SwinJSCC work, where $R_t$ represents the ratio between the number of transmitted symbols and the number of input pixels. The generated symbols are then appended to the transmission buffer and wait for transmission.

\subsubsection{Reward Function}
An image is considered successfully transmitted if its reconstruction quality meets predefined thresholds $\Gamma_{\text{PSNR}}$ and $\Gamma_{\text{MS-SSIM}}$. The immediate reward is defined as:
\begin{equation}
\label{eq:r_t}
r_t = 
\begin{cases}
1, & \text{if } \text{PSNR}_t \geq \Gamma_{\text{PSNR}} \text{ and } \text{MS-SSIM}_t \geq \Gamma_{\text{MS-SSIM}}, \\
0, & \text{otherwise}.
\end{cases}
\end{equation}
To enforce system constraints and encourage efficient buffer utilization, we augment the reward with two queue-related penalty terms.

First, when the queue length exceeds the threshold $Q_{\text{th}}$, we apply a linear penalty that grows with the degree of overflow:
\[
p_{\text{over}}(t) = \lambda_{\text{over}} \cdot \max\left(0, \frac{Q_t - Q_{\text{th}}}{Q_{\max} - Q_{\text{th}}}\right),
\]
where $\lambda_{\text{over}}$ is a coefficient controlling the severity of the penalty, and the fraction normalizes the excess to $[0,1]$. This term discourages the agent from allowing the queue to approach saturation.

Second, to prevent the queue from staying unnecessarily empty (which may indicate underutilization of the downlink), we introduce a small penalty when the normalized queue length is below a lower threshold $Q_{\text{low}}$:
\[
p_{\text{under}}(t) = \lambda_{\text{under}} \cdot \mathbf{1}_{\{Q_t \le Q_{\text{low}}\}},
\]
where $\lambda_{\text{under}}$ is a small positive coefficient. This term gently encourages the agent to maintain a minimal buffer occupancy, avoiding excessive idle periods.

Additionally, a discard penalty remains for images dropped due to queue overflow:
\[
p_{\text{drop}}(t) = \lambda_d \cdot d_t,
\]
where $d_t$ is the number of discarded images in time slot $t$.

The overall augmented reward is then:
\[
\tilde{r}_t = r_t - p_{\text{over}}(t) - p_{\text{under}}(t) - p_{\text{drop}}(t),
\]
where $r_t = \mathbf{1}_{\{\text{PSNR}_t \ge \Gamma_{\text{PSNR}} \land \text{MS-SSIM}_t \ge \Gamma_{\text{MS-SSIM}}\}}$ is the success indicator for transmitted images (defined in Eq.~(\ref{eq:r_t})).



\subsubsection{Optimization Objective}
    
We formulate the problem as a finite-horizon, discrete-time Markov decision 
process (MDP) defined by the tuple $(\mathcal{S}, \mathcal{A}, P, \tilde{r}, T)$, 
where $\mathcal{S}$ is the state space; $\mathcal{A}$ is the action space; 
$P$ is the transition probability, defined implicitly by the system dynamics (i.e., the evolution of SNR, elevation angle and buffer length); $\tilde{r}$ is the reward function, and $T$ is the horizon. The goal is to find an optimal policy $\pi^*$ that maximizes the expected cumulative reward, which corresponds to transmitting as many qualified images as possible within each visibility window:

\begin{equation}
\pi^* = \arg\max_{\pi} \; \mathbb{E}_{\pi} \left[ \sum_{t=1}^{T} \tilde{r}_t \right],
\end{equation}

subject to:
\begin{align}
& \text{PSNR}_t \geq \Gamma_{\text{PSNR}}, \quad \forall t, \\
& \text{MS-SSIM}_t \geq \Gamma_{\text{MS-SSIM}}, \quad \forall t, \\
& Q_t \leq Q_{\max}, \quad \forall t, \\
& \sum_{t=1}^{T} T_{\text{tx}}(R_t) \leq T_{\text{win}}.
\end{align}

\section{Proposed Approach}
\label{sec:approach}
\begin{figure*}
    \centering
    \includegraphics[width=0.90\linewidth]{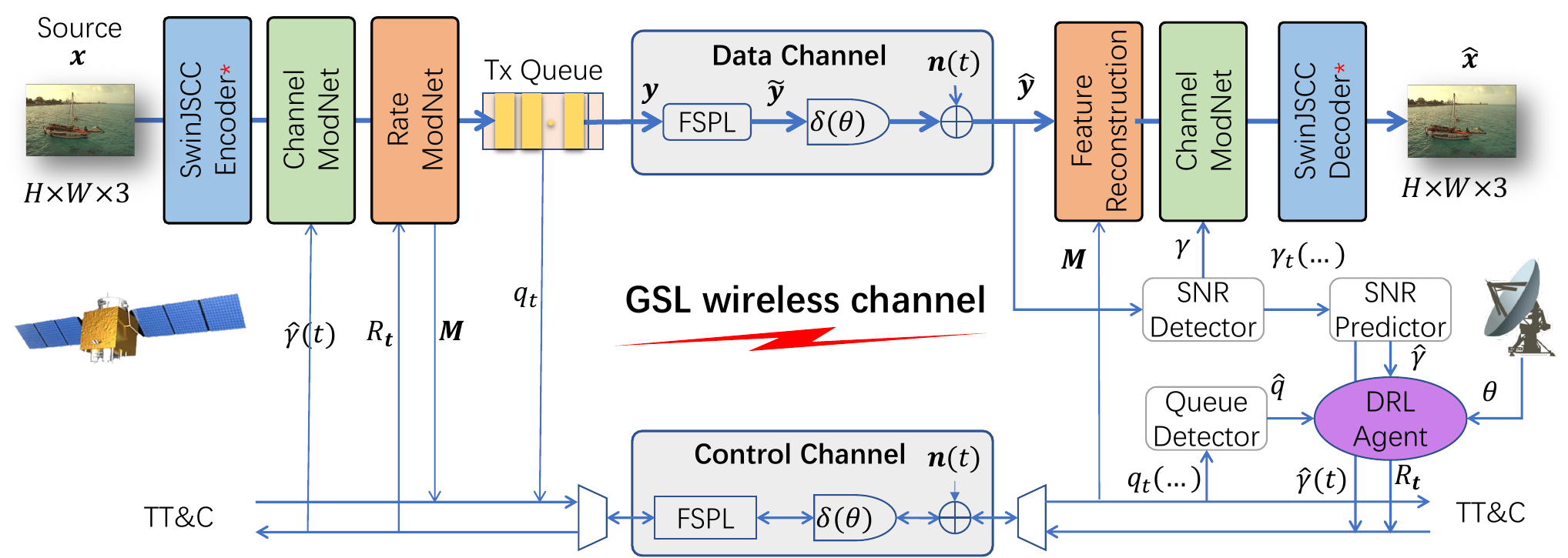}
    \caption{The overall architecture of GSL-SwinJSCC for image transfer. }
    \label{fig:arch}
\end{figure*}

The overall architecture of our proposed GSL-SwinJSCC is illustrated in Fig.~\ref{fig:arch}. 
The images captured by the satellite are first processed by a SwinJSCC encoder, which extracts semantic features. The encoder is fine-tuned to accommodate the GSL SNR range (1–38 dB), ensuring robust performance across the full overpass. After channel and rate modulation, the encoded features are placed into a transmission queue. The queue is introduced to accommodate the limited on-board storage capacity and the bandwidth bottleneck of GSL. The transmission queue has a maximum symbol capacity of $Q_{\max}$ and transmits the features to the ground receiver at a constant symbol rate $R_s$ over the GSL data channel. At the ground side, the received signals undergo feature reconstruction and image restoration, leveraging both the SNR estimated by an SNR detector and the mask vector used for rate modulation, which is received via the control channel.

To fully utilize the downlink bandwidth and cope with the long delay and dynamic nature of the GSL, we introduce a DRL agent along with an SNR predictor. The predictor forecasts the future SNR based on GSL link budget. The predicted SNR serves as the input to the on-board channel modulation module, and together with the current queue length and the ground elevation angle, forms the system state variables. Based on this state, the DRL agent determines an optimized compression ratio $R$, which serves as the target rate for the rate modulation network. These control parameters are reliably transmitted via the satellite's tracking, telemetry, and command (TT\&C) system.

\subsection{Prediction-Aligned Control}
Although the LEO feeder link is well approximated by an AWGN model with deterministic SNR variations, the physical propagation delay and the uplink command transmission introduce a non-negligible control loop latency of approximately $10$–$15$ ms. During this interval, the satellite elevation angle and consequently the SNR $\gamma_t$ can change significantly, especially at low elevation angles where the SNR slope is steep. Moreover, the satellite's transmission buffer occupancy $Q_t$ evolves dynamically in response to previous compression decisions and the constant symbol rate of the physical layer. A decision made based on the current SNR $\gamma_t$ and buffer length $Q_t$ may therefore become outdated by the time the corresponding image is actually transmitted, leading to suboptimal compression ratio selection, buffer overflow, or quality degradation.

To address this mismatch, the reinforcement learning agent is deployed at the ground gateway to leverage its abundant computational resources. The architecture follows a \emph{predict–decide–execute} paradigm that compensates for the inherent control loop latency in LEO satellite links.

\subsubsection{Downlink Telemetry and Prediction}
The satellite periodically samples its transmission buffer occupancy $Q_t$ and encapsulates this information into CCSDS-compliant telemetry frames transmitted to the ground gateway. Concurrently, the ground gateway estimates the downlink SNR $\gamma_t$ from the received signal using embedded pilots or preamble sequences. The gateway maintains a historical buffer of both the telemetry (queue length) and the locally estimated SNR, capturing the temporal evolution of the system state.

A SNR predictor operates at the gateway using a lightweight, precomputed mechanism based directly on the link budget. Specifically, it leverages the deterministic relationship between elevation angle and SNR established by the link budget formula (Eq.~(2)). Given the current elevation angle $\theta_t$ and the satellite orbital dynamics, the elevation angle at the next decision step $\theta_{t+1}$ is known. The corresponding SNR $\hat{\gamma}_{t+1}$ is then obtained by evaluating the link budget at $\theta_{t+1}$. Note that the control loop latency (approximately $10$–$15$ ms) is negligible compared to the decision interval (e.g., $5$ seconds); therefore, the prediction for the next decision step effectively compensates for the delay without requiring an explicit $\Delta t$ parameter.

\subsubsection{RL Decision Making}
At each decision epoch, the RL agent constructs a state vector using the predicted future values according to Eq.~(\ref{eqn:state}). The agent then selects an action $a_t = R_t \in \mathcal{R}$, i.e., the compression ratio for the next image.

The selected compression ratio $R_t$ and the SNR value (either the current $\gamma_t$ or the predicted $\hat{\gamma}_{t+1}$) are formatted into an uplink command and transmitted to the satellite. Upon reception, the satellite first updates the \emph{Channel ModNet} with the received SNR to adapt the encoding strength, followed by the \emph{Rate ModNet} with $R_t$ to control the number of symbols per image. The choice of feeding either $\gamma_t$ or $\hat{\gamma}_{t+1}$ to the Channel ModNet is examined in the ablation study.

This \emph{ground-based measurement, prediction, and decision} architecture ensures that the satellite remains lightweight while benefiting from sophisticated adaptive control. The total control loop latency is effectively compensated by the SNR predictor, which aligns the decisions with the future channel condition at the time of transmission to a certain degree.

\subsection{Choice of Reinforcement Learning Algorithm}

Given that our action space $\mathcal{A} = \mathcal{R}$ is discrete with $K = 5$ possible compression ratios, a value-based deep Q-learning approach is a natural fit. We adopt the standard Deep Q-Network (DQN) as the core learning algorithm for its simplicity and proven effectiveness in discrete control tasks.

We also experimented with more advanced variants, including Double DQN (DDQN) and Dueling DQN, to assess whether they could bring measurable improvements in our specific setting. However, extensive simulation results indicated that neither DDQN nor Dueling DQN consistently outperformed the standard DQN under our LEO satellite scenario. The performance gains, if any, were marginal and did not justify the added implementation complexity.

Consequently, we retain the standard DQN as the algorithm of choice for this work. It is worth emphasizing that the main contribution of this paper is not the invention of a new reinforcement learning algorithm, but rather the integrated framework that combines a symbol-level transmission queue, SNR prediction, and ground-based closed-loop rate adaptation. The DQN serves as a capable and lightweight decision engine within this framework, and more sophisticated RL algorithms can be seamlessly substituted in future extensions if needed.

\section{Simulation Results and Analysis}
\label{sec:result}

\subsection{Experiment Setup}
\begin{table}[!t]
\caption{Parameters for the satellite-to-ground feeder links used for generation of SNR data\label{tab:gsl_parameters}}
\centering
    \begin{tabular}{|c||c|}
        \hline
        Carrier Frequency (GHz) & 20\\
        \hline
        Orbit Altitude (km) & 900\\
        \hline
        Satellite EIRP (dBW) & 35\\
        \hline
        Ground Station G/T (dB/K) & 25\\
        \hline
        Receiver Noise Bandwidth (MHz) & 100\\
        \hline
        Extra loss (dB) & 2.5\\
        \hline
        Signal Bandwidth (MHz) & 200 \\
    \hline
    \end{tabular}
\end{table}
\subsubsection{Simulation Environment} 
We evaluate our proposed RL-based rate adaptation framework on a simulated LEO satellite-to-ground link. The key system parameters follow those listed in Table~\ref{tab:gsl_parameters}. For all subsequent simulations involving SNR sweeps, we directly employ the analytical link budget formula (Eq.~(\ref{eqn:snr})), which maps the elevation angle $\theta$ to SNR values. Specifically, as $\theta$ increases from $20^\circ$ to $90^\circ$ and then decreases back to $20^\circ$ over a complete visibility window, the corresponding SNR varies between approximately $26$ dB and $34$ dB. The decision interval is set to $T_d = 5$ seconds, and the encoder processes $M = 12$ images per decision interval. Each image has a resolution of $768 \times 512$ pixels (Kodak24) with $C_{\text{in}} = 3$ channels (RGB). The SwinJSCC model used in all experiments is the fine-tuned version described in Section~V, which supports SNR ranging from $1$ dB to $38$ dB. The quality thresholds are set to $\Gamma_{\text{PSNR}} = 32$ dB and $\Gamma_{\text{MS-SSIM}} = 0.94$ by default. 

\subsubsection{Compared Algorithms}
We compare our proposed method against the following baseline policies:
\begin{itemize}

    \item \textbf{Fixed Maximum Rate} (\texttt{max\_rate}): The encoder always uses the maximum allowed number of channels ($C = 192$) to transmit semantic features for every image.
    \item \textbf{Fixed Medium Rate} (\texttt{mid\_rate}): The encoder always uses the medium allowed number of channels ($C = 96$) for every image.
    \item \textbf{Fixed Minimum Rate} (\texttt{min\_rate}): The encoder always uses the minimum allowed number of channels ($C = 32$) for every image.
    \item \textbf{DQN} (\texttt{rl\_dqn}): A standard Deep Q-Network for rate adaptation.

\end{itemize}
\subsubsection{Evaluation Metrics}
We adopt the following metrics to evaluate the performance of different rate adaptation policies.
\begin{itemize}
    \item \textbf{Qualified Frames ($Q_{\text{qual}}$)}: The number of successfully transmitted images that meet both quality thresholds, i.e., $\text{PSNR} \geq \Gamma_{\text{PSNR}}$ and $\text{MS-SSIM} \geq \Gamma_{\text{MS-SSIM}}$, accumulated over a complete satellite overpass. This is the primary measure of effective throughput.

    \item \textbf{Forwarding Efficiency ($\eta_1$)}: Defined as $\eta_1 = Q_{\text{qual}} / F$, where $F$ denotes the number of frames that leave the transmission queue and are transmitted over the wireless channel (i.e., forwarded frames). This metric captures the proportion of transmitted frames that ultimately become qualified. It penalizes policies that waste on-board computational resources, storage, and transmission power on frames that fail to meet the quality requirements.
\end{itemize}

\subsection{Training}
\subsubsection{Fine-tune of SwinJSCC}
The pre-trained SwinJSCC model is trained on AWGN channels with SNR values ranging from 1 to 13 dB. However, our link-budget based GSL SNR sequence exhibits a wider dynamic range: the SNR reaches approximately 34 dB at zenith under clear-sky conditions, significantly exceeding the upper bound of the original pre-trained model.

To address this domain gap, we fine-tune the pre-trained SwinJSCC model on our GSL-specific SNR distribution. Specifically, we adopt the versatile version equipped with both Channel Adaptation (SA) and Rate Adaptation (RA) modules. The pre-trained weights of the Swin Transformer backbone are preserved as initialization, while the model is further trained on simulated channel conditions covering the full SNR range observed in our GSL scenarios. Notably, the primary focus of fine-tuning is on the SNR values corresponding to elevation angles above $20^\circ$, which fall within the range of approximately $30.58$–$38.0$ dB. This fine-tuning strategy enables the model to adapt its encoding and decoding behavior to the realistic SNR dynamics of LEO satellite links, thereby improving reconstruction quality across the entire overpass, with particular emphasis on the high-SNR regimes encountered at moderate to high elevation angles.

\begin{figure}[htbp]
  \centering
  \subfloat[PSNR]{
    \includegraphics[width=0.45\linewidth]{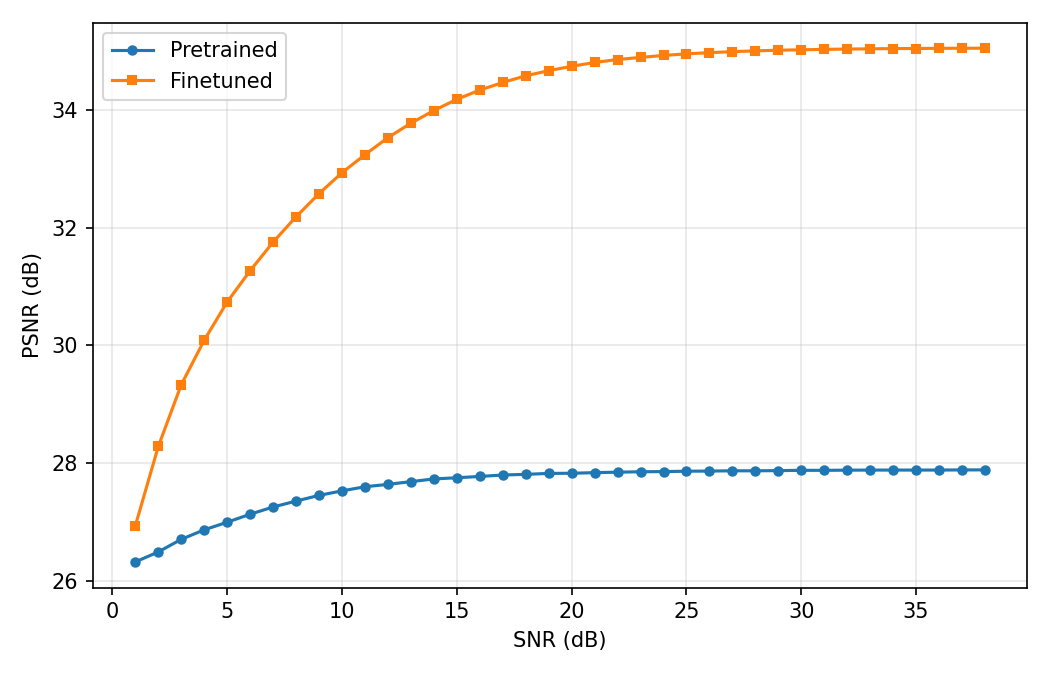}
    \label{fig:ft_psnr}
  }\hfill
  \subfloat[MS-SSIM]{
    \includegraphics[width=0.45\linewidth]{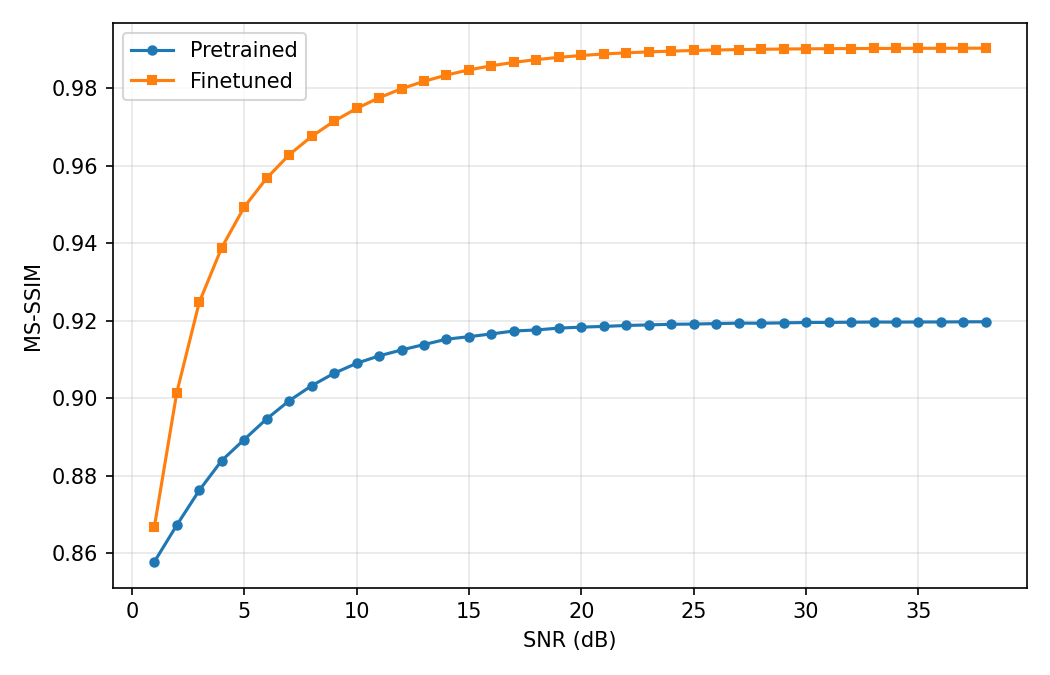}
    \label{fig:ft_mssin}
  }
  \caption{Results of model finetuning.}
  \label{fig:finetune}
\end{figure}

As shown in Fig.~\ref{fig:finetune}, the fine-tuned model achieves significantly better reconstruction quality across the entire SNR range (1--38 dB) in terms of both PSNR and MS-SSIM compared to the base model when a moderate channel dimension ($C = 96$) is adopted. Moreover, when the SNR exceeds $5$ dB, the fine-tuned model yields PSNR above $30.8$ dB and MS-SSIM above $0.95$.

\subsubsection{DRL Training}
During training, the RL agent operates in a simulated environment where the 
original images are readily available at the gateway. This allows immediate 
computation of PSNR and MS-SSIM after each transmission, providing instant 
reward signals for policy optimization.

\subsection{Simulation Results}
\begin{table*}[!t]
\caption{Comparison of different policy on total qualified frames received. \label{tab:policy}}
\centering
    \begin{tabular}{|c||c|c|c|c|c|c|}
        \hline
        Policy & Qualified & Forwarded & Dropped & Mean Ch\_num & Mean CBR & Qual/Fwd (\%) \\
        \hline
        \textbf{rl\_dqn}   & \textbf{524} & \textit{564} & 0   & \textit{115.84}  & \textit{0.076241} & \textit{92.91}\\
        \hline
        \textbf{max\_rate} & 343 & 343 & 228 & \textbf{192.00} & \textbf{0.12500}  & \textbf{100.00}\\
        \hline
        \textbf{mid\_rate} & \textit{499} & \textbf{588} & 0   & 96.00  & 0.062500 & 84.86\\ 
        \hline
        \textbf{min\_rate} & 250 & \textbf{588} & 0   & 32.00  & 0.020833 & 42.52\\ 
    \hline
    \end{tabular}
\end{table*}
\subsubsection{SNR sweep and rate adjust} 
\begin{figure}[htbp]
  \centering
  \subfloat[SNR sweep]{
    \includegraphics[width=0.85\linewidth]{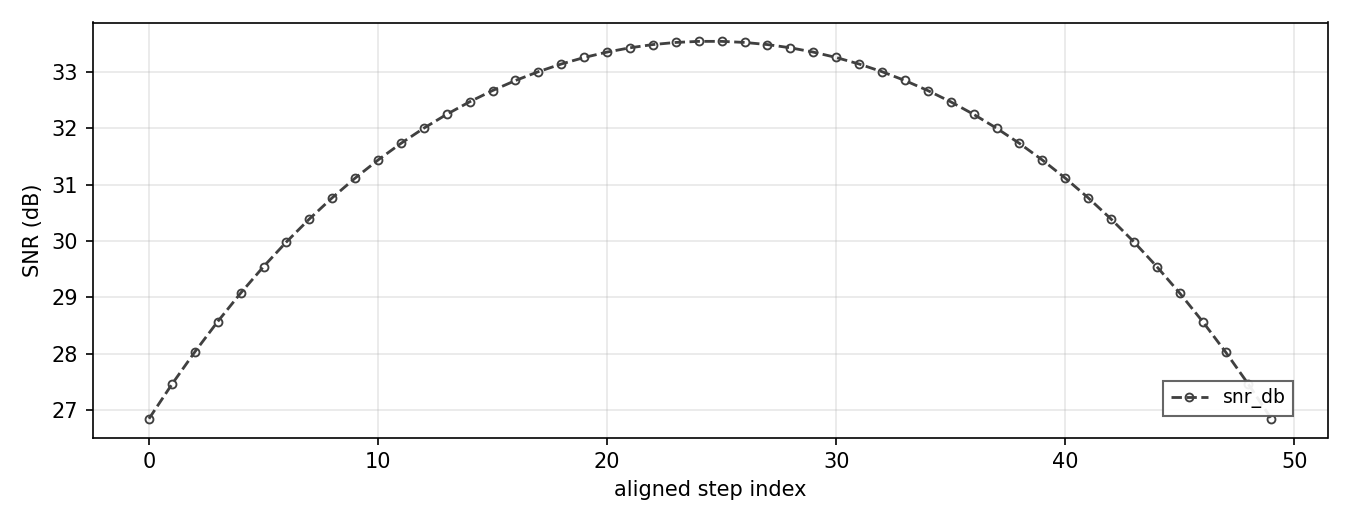}
    \label{fig:snr_sweep}
  }\hfill
  \subfloat[Target rate adjust]{
    \includegraphics[width=0.85\linewidth]{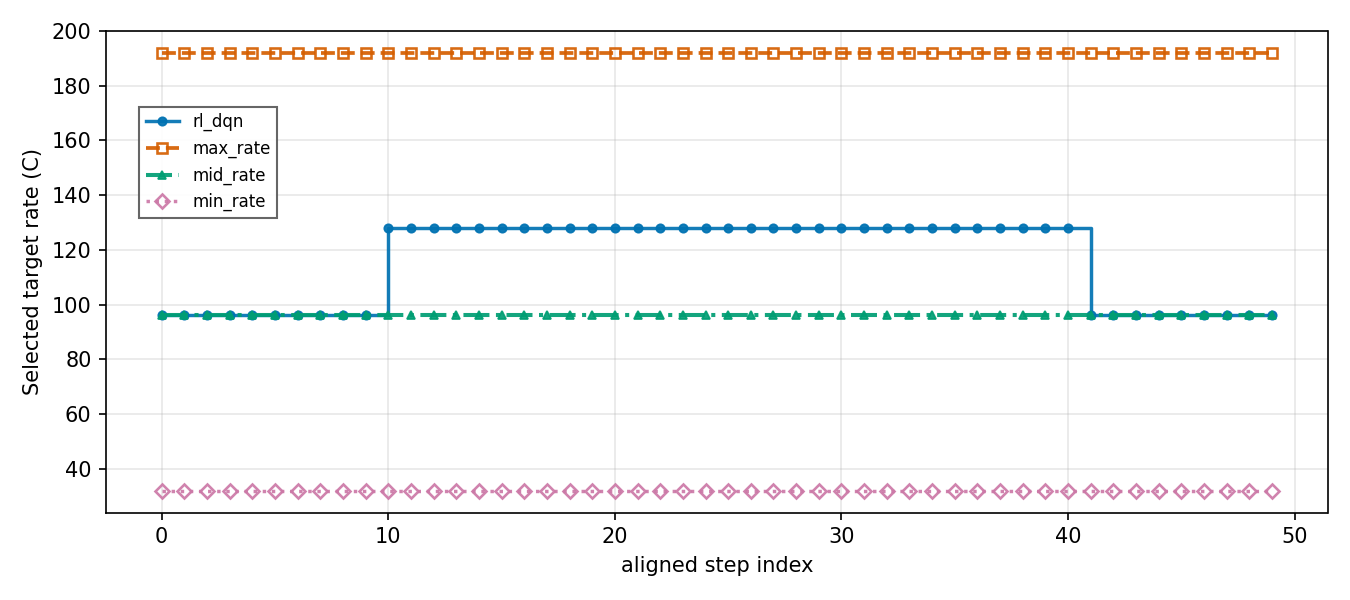}
    \label{fig:rate_adjust}
  }
  \hfill
  \subfloat[Tx\_buffer occupancy]{
    \includegraphics[width=0.85\linewidth]{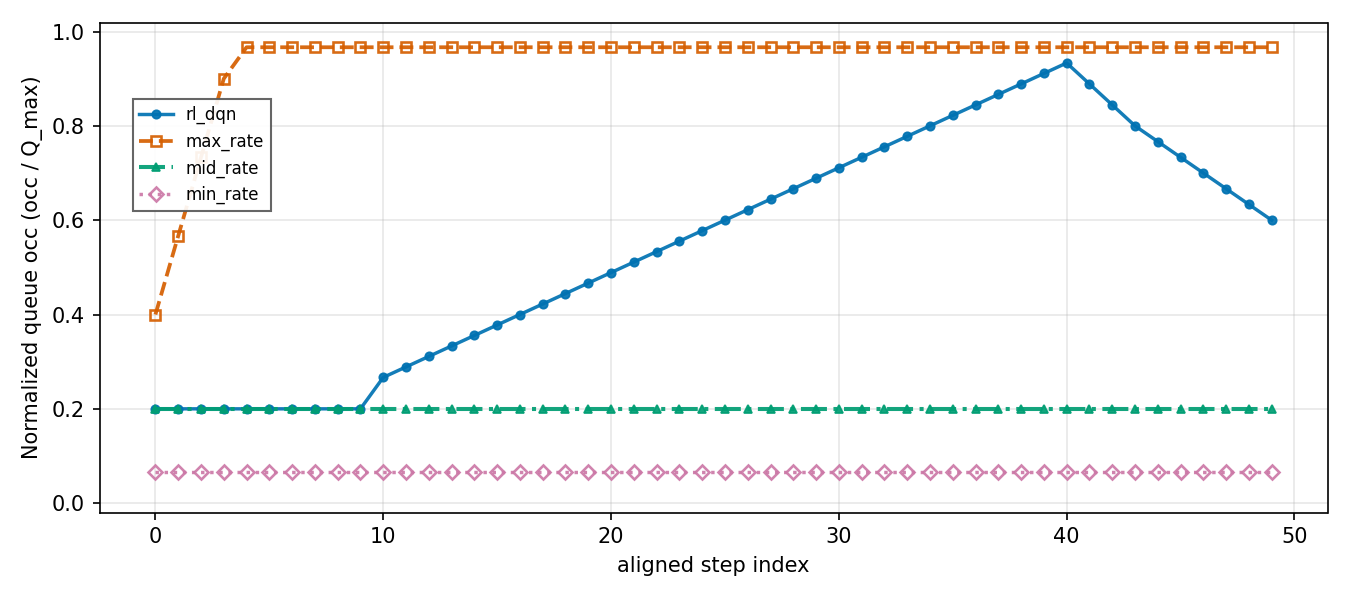}
    \label{fig:occb_qnorm}
  }
  \hfill
  \subfloat[Qualified Frames per step]{
    \includegraphics[width=0.85\linewidth]{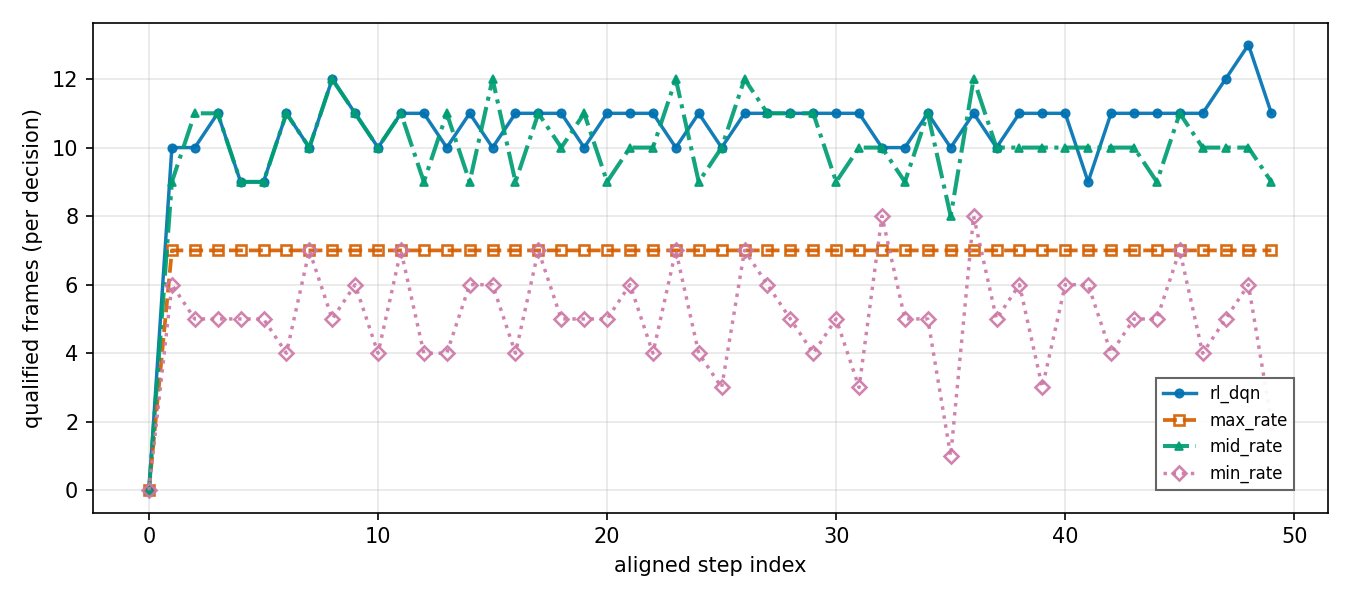}
    \label{fig:qual_frames}
  }
  \caption{Process of SNR sweep and adjust of target rate at $\Gamma_{\text{PSNR}} = 32$ dB and $\Gamma_{\text{MS-SSIM}} = 0.94$.}
  \label{fig:snr_sweep_rate_qual}
\end{figure}

To emulate a full satellite overpass, the elevation angle $\theta$ is swept from $0^\circ$ up to $90^\circ$ and then back to $0^\circ$. The instantaneous SNR, obtained from the link budget calculation, exhibits a characteristic low–high–low trajectory over the visibility window, as shown in Fig.~\ref{fig:snr_sweep}. The evolution of the target rate under different policies is shown in Fig.~\ref{fig:rate_adjust}. The fixed-rate baselines, \texttt{min\_rate}, \texttt{mid\_rate}, and \texttt{max\_rate}, adopt constant channel numbers $C = 32$, $96$, and $192$, respectively. In contrast, \texttt{rl\_dqn} starts at $96.0$, then after step $10$ jumps to $128.0$, and later returns to $96.0$ at step $41$, illustrating adaptive rate adjustment based on channel conditions and queue status. Figure~\ref{fig:occb_qnorm} shows the transmission buffer occupancy of different policies during the entire SNR sweep. The \texttt{min\_rate} and \texttt{mid\_rate} policies maintain average occupancies of $6.7\%$ and $20\%$, respectively. In contrast, \texttt{max\_rate} saturates the buffer as early as step $5$, leading to packet losses, with an average occupancy as high as $96.7\%$. For the proposed \texttt{rl\_dqn} policy, the buffer occupancy stays at around $20\%$ during the first $9$ steps when the target rate is $96$. From steps $10$ to $40$, the agent increases the target rate, causing the buffer utilization to rise steadily and peak at $91.2\%$. After the target rate decreases, the occupancy gradually drops to $62.5\%$. Notably, the proposed policy achieves high buffer utilization while incurring no packet loss throughout the entire process. Figure~\ref{fig:qual_frames} and Table~\ref{tab:policy} present the per-step and overall performance of different policies. \texttt{max\_rate} achieves $7$ qualified frames per step but drops $228$ frames in total due to buffer overflow, resulting in only $343$ total qualified frames. \texttt{min\_rate} avoids drops but only $42.52\%$ of its forwarded frames are qualified ($250$ frames). Both \texttt{mid\_rate} and \texttt{rl\_dqn} have no drops, achieving qualified ratios of $84.86\%$ and $92.91\%$, respectively. Notably, \texttt{rl\_dqn} exploits favorable channel conditions by adopting a higher target rate, thereby obtaining more qualified frames than \texttt{mid\_rate}.
\subsubsection{PSNR and MS-SSIM distributions of reconstrcuted images} 
\begin{figure*}[htbp]
  \centering
  \subfloat[]{
    \includegraphics[width=0.23\linewidth]{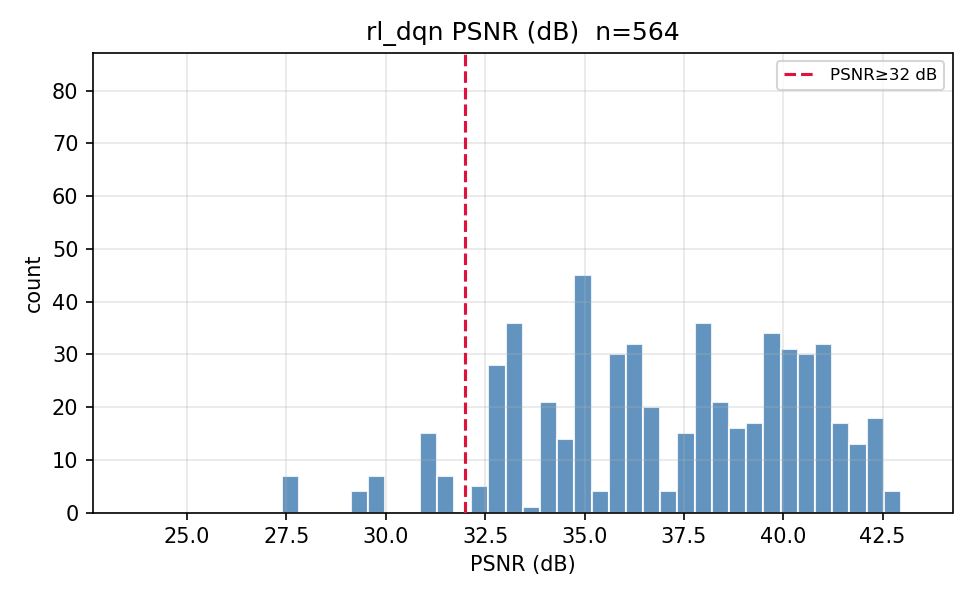}
    \label{fig:rl_dqn_psnr_hist}
  }\hfill
  \subfloat[]{
    \includegraphics[width=0.23\linewidth]{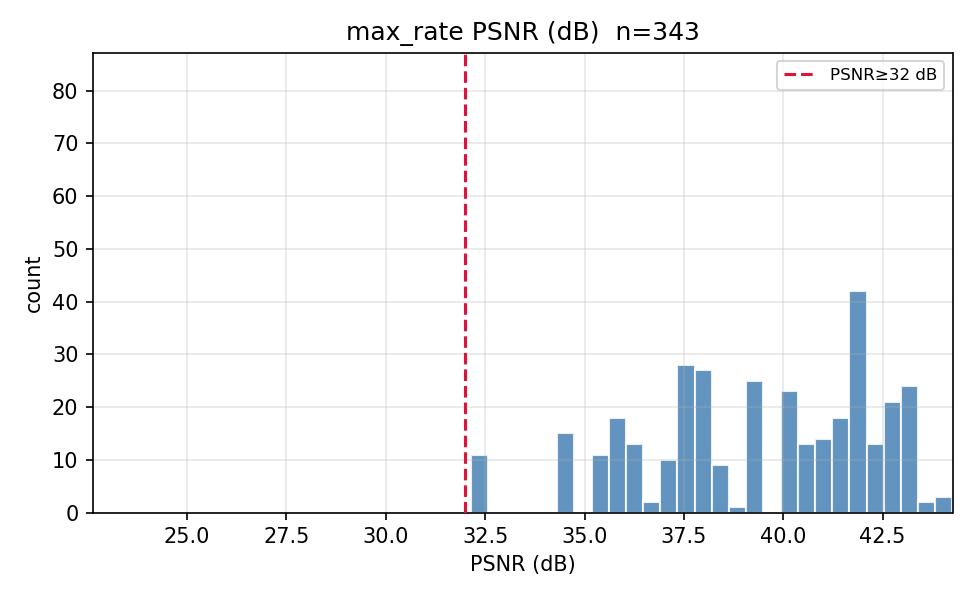}
    \label{fig:max_rate_psnr_hist}
  }
  \hfill
  \subfloat[]{
    \includegraphics[width=0.23\linewidth]{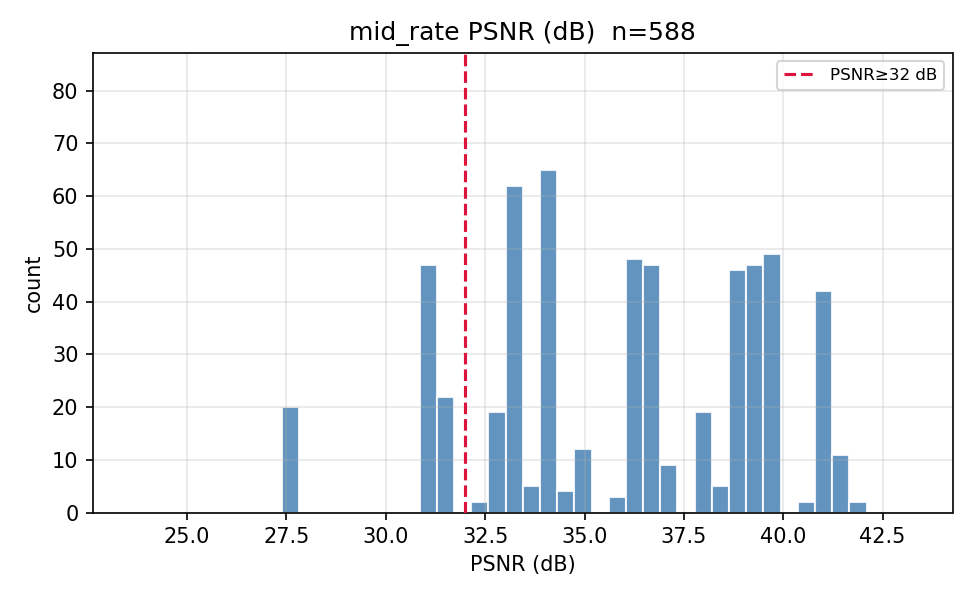}
    \label{fig:mid_rate_psnr_hist}
  }\hfill
  \subfloat[]{
    \includegraphics[width=0.23\linewidth]{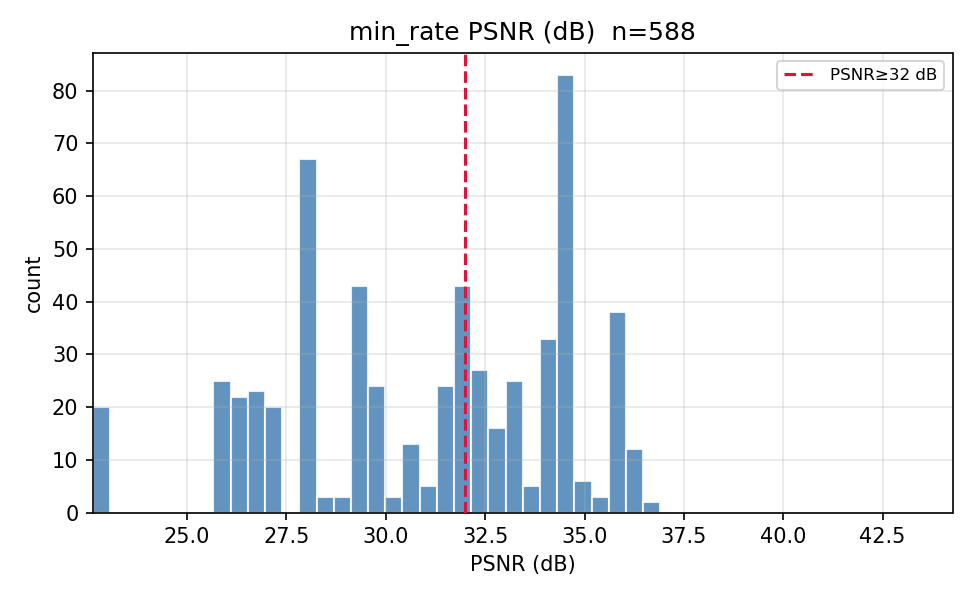}
    \label{fig:min_rate_psnr_hist}
  }
  \vfill
  \subfloat[]{
    \includegraphics[width=0.23\linewidth]{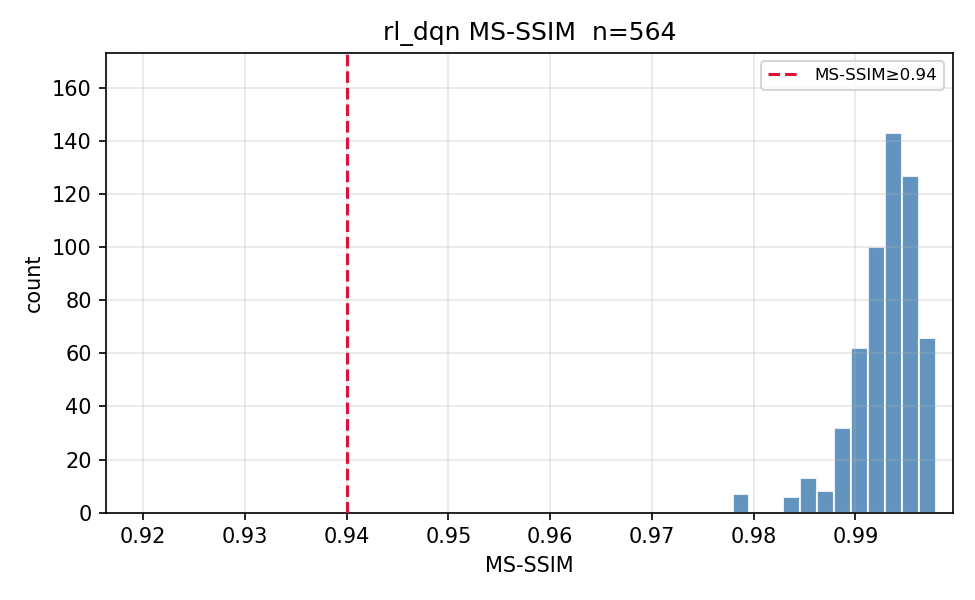}
    \label{fig:rl_dqn_msssim_hist}
  }\hfill
  \subfloat[]{
    \includegraphics[width=0.23\linewidth]{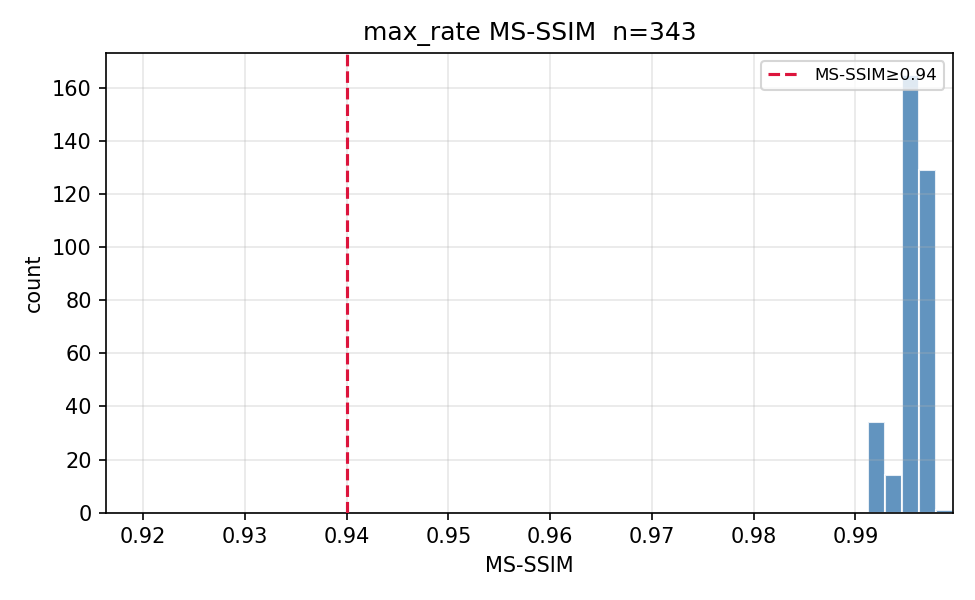}
    \label{fig:max_rate_msssim_hist}
  }
  \hfill
  \subfloat[]{
    \includegraphics[width=0.23\linewidth]{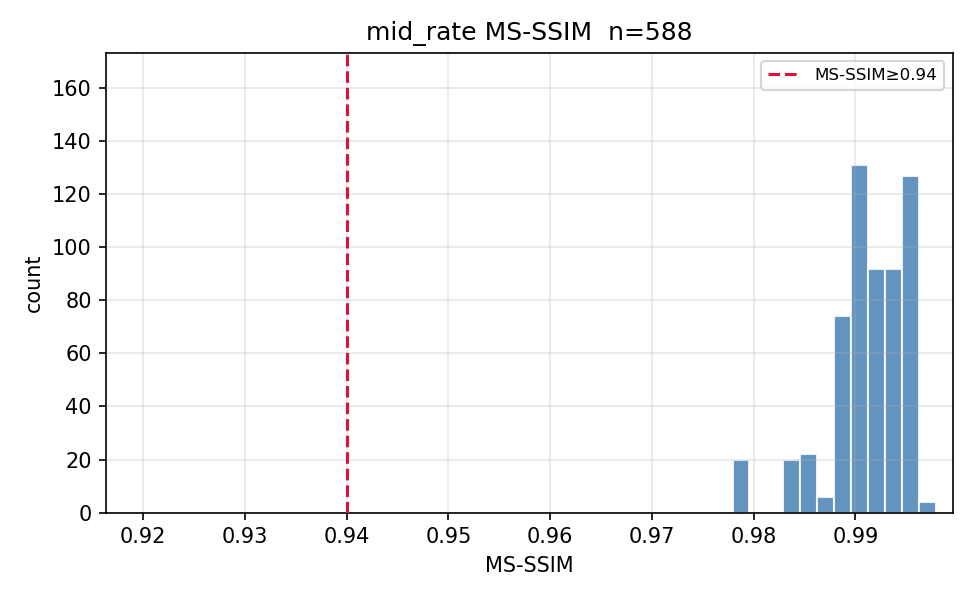}
    \label{fig:mid_rate_msssim_hist}
  }\hfill
  \subfloat[]{
    \includegraphics[width=0.23\linewidth]{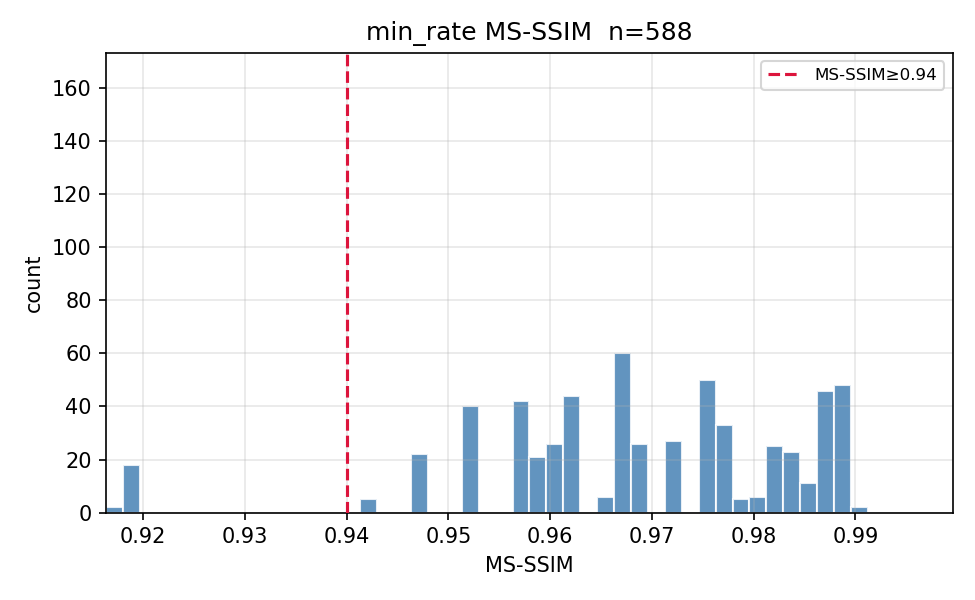}
    \label{fig:min_rate_msssim_hist}
  }
  \caption{Distributions of PSNR (a–d) and MS-SSIM (e–h) for reconstructed images. 
         (a)–(d) show PSNR for \texttt{rl\_dqn}, \texttt{max\_rate}, \texttt{mid\_rate}, 
         and \texttt{min\_rate}, respectively; (e)–(h) show the MS-SSIM distributions 
         for the same four methods.}
  \label{fig:psnr_msssim_dist}
\end{figure*}

Figure~\ref{fig:psnr_msssim_dist} illustrates the distributions of PSNR and MS-SSIM for the reconstructed images, with the quality thresholds set to $\Gamma_{\text{PSNR}} = 32$ dB and $\Gamma_{\text{MS-SSIM}} = 0.94$, as indicated by the red dashed lines in the figure. From panels (a)–(d), it can be observed that, except for the \texttt{max\_rate} policy, all other three strategies produce some PSNR values below the threshold, with \texttt{min\_rate} having the most such instances. This result is consistent with expectations. Panels (e)–(h) show that, except for \texttt{min\_rate}, the MS-SSIM values of the other three policies are consistently far above the threshold, typically around $0.99$. Only \texttt{min\_rate} occasionally drops to approximately $0.92$. Therefore, unqualified frames are primarily caused by insufficient PSNR rather than MS-SSIM degradation.
\subsubsection{Viuallization}
\begin{figure}[htbp]
  \centering
  \subfloat[\texttt{rl\_dqn}: psnr=27.5748]{
    \includegraphics[width=0.95\linewidth]{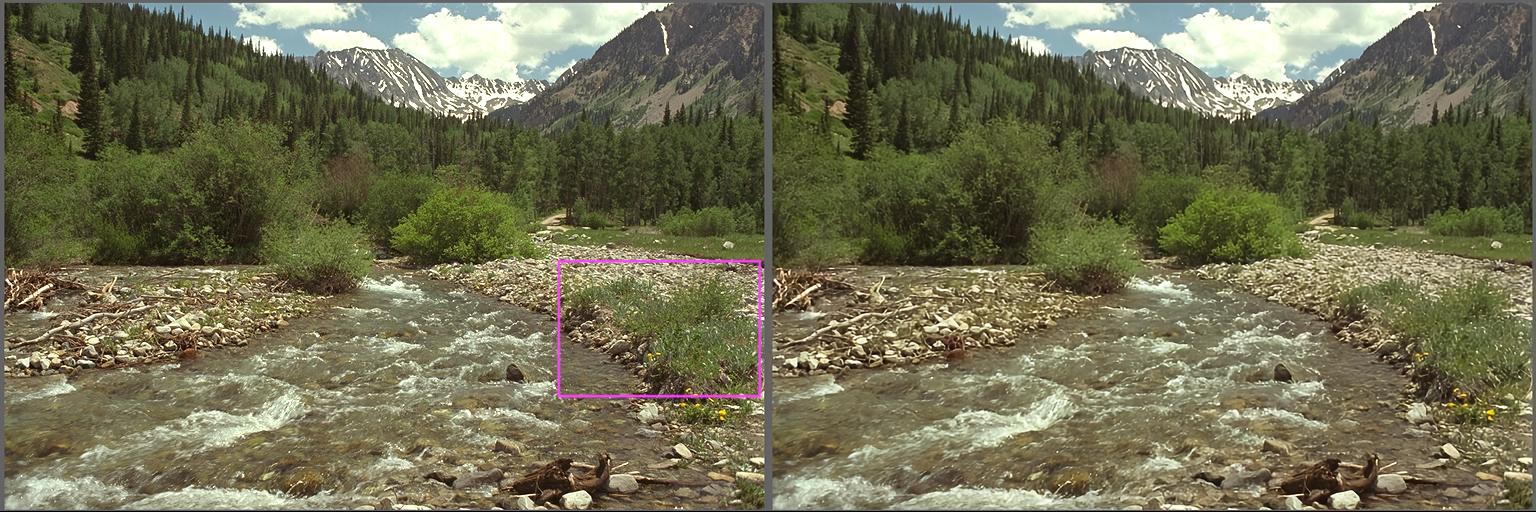}
    \label{fig:dqn_psnr_min}
  }\hfill
  \subfloat[\texttt{max\_rate}: psnr=32.2922]{
    \includegraphics[width=0.95\linewidth]{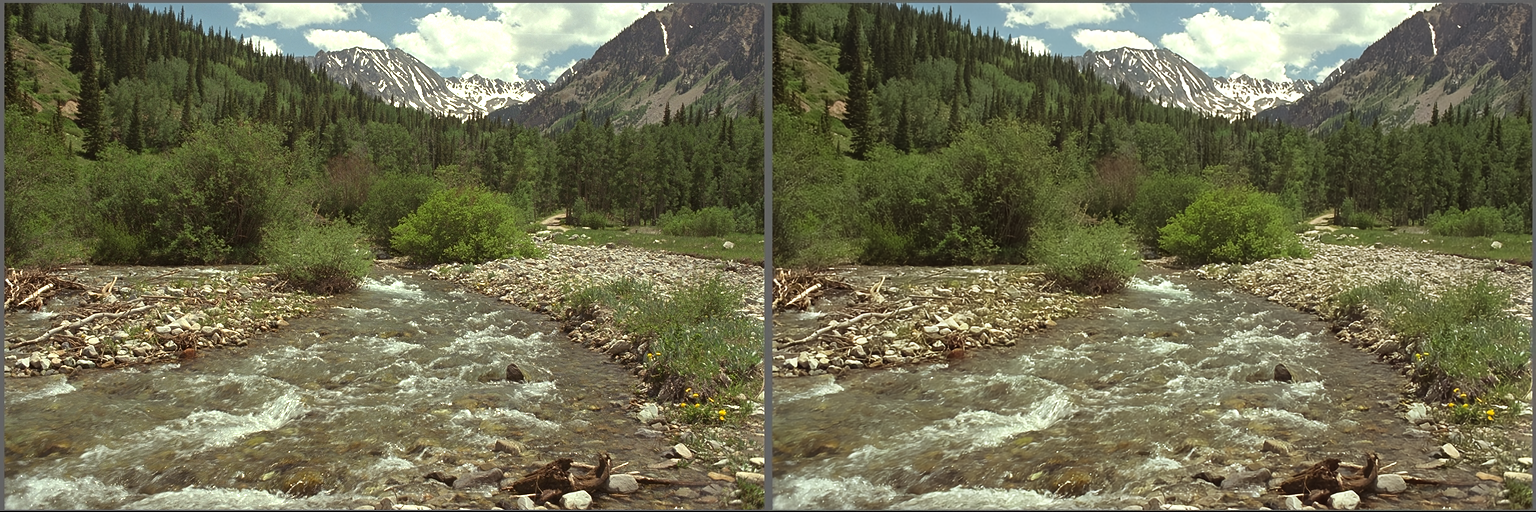}
    \label{fig:max_rate_psnr_min}
  }
  \hfill
  \subfloat[\texttt{mid\_rate}: psnr=27.5639]{
    \includegraphics[width=0.95\linewidth]{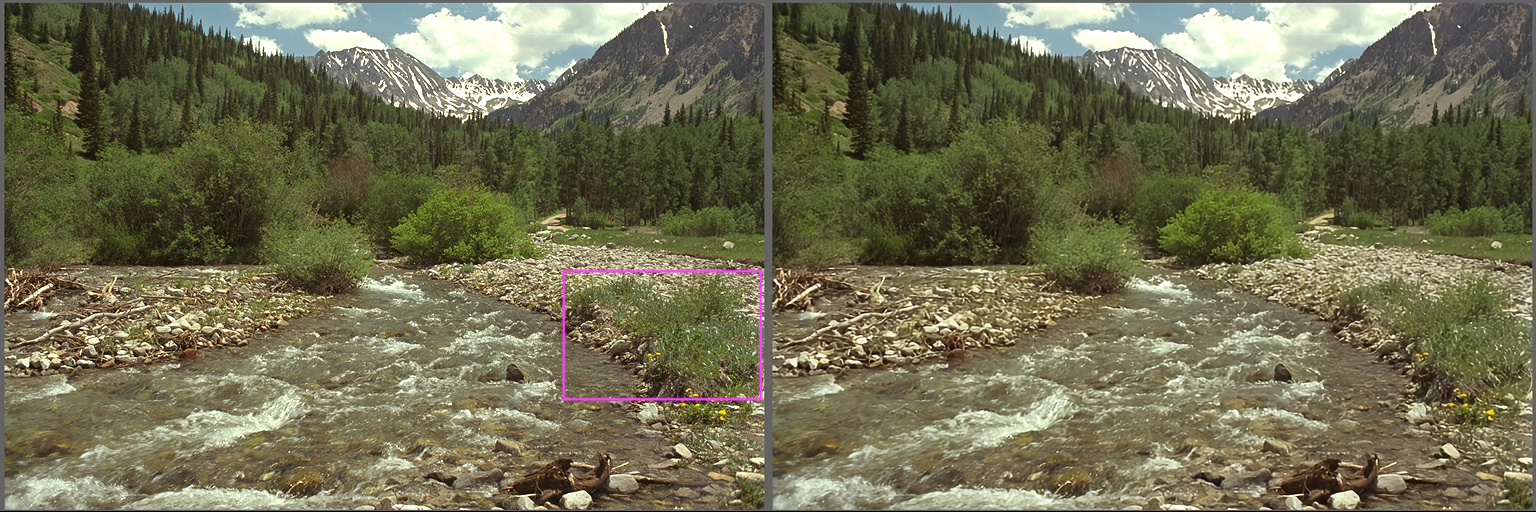}
    \label{fig:mid_rate_psnr_min}
  }
  \hfill
  \subfloat[\texttt{min\_rate}: psnr=23.0531]{
    \includegraphics[width=0.95\linewidth]{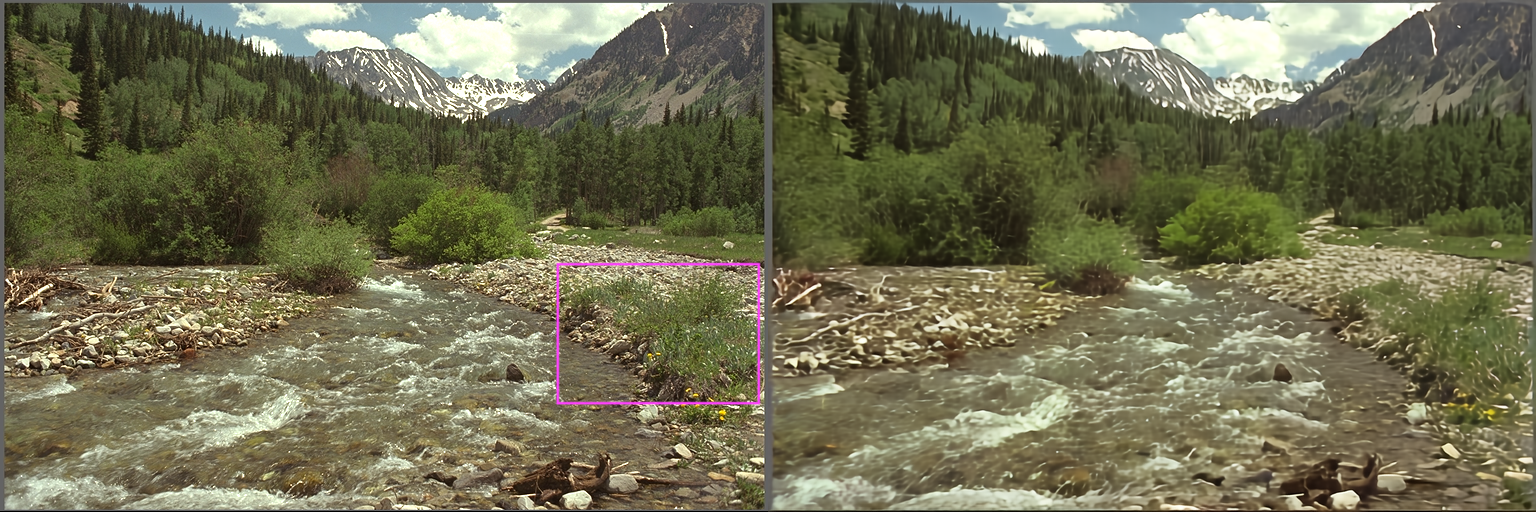}
    \label{fig:min_rate_psnr_min}
  }
  \caption{Comparison of reconstruction quality for the lowest-PSNR image per policy. Left: original; right: reconstructed.}
  \label{fig:psnr_min_orig_recon}
\end{figure}

Figure~\ref{fig:psnr_min_orig_recon} presents a side-by-side comparison of the original image and the reconstructed image with the \textit{lowest} PSNR under each policy. The left column shows the original image, while the right column shows the reconstructed one. As expected, \texttt{min\_rate} yields the poorest reconstruction quality (PSNR: $23.05$ dB), \texttt{max\_rate} achieves the best (PSNR: $32.29$ dB), while both \texttt{mid\_rate} and \texttt{rl\_dqn} produce intermediate quality with PSNR values of $27.57$ dB and $27.56$ dB, respectively.
\subsubsection{Ablation Study: Impact of SNR Prediction}
\begin{figure}[htbp]
  \centering
  \subfloat[Target rate adjust]{
    \includegraphics[width=0.85\linewidth]{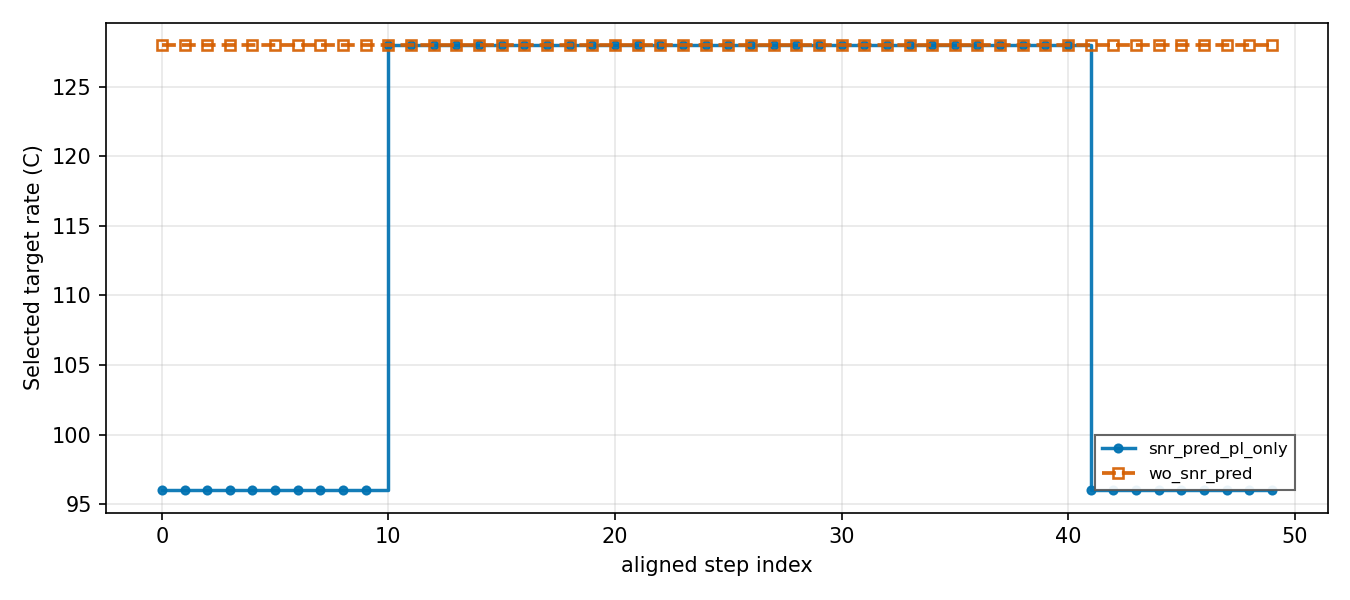}
    \label{fig:ab_rate_adjust}
  }
  \hfill
  \subfloat[Tx\_buffer occupancy]{
    \includegraphics[width=0.85\linewidth]{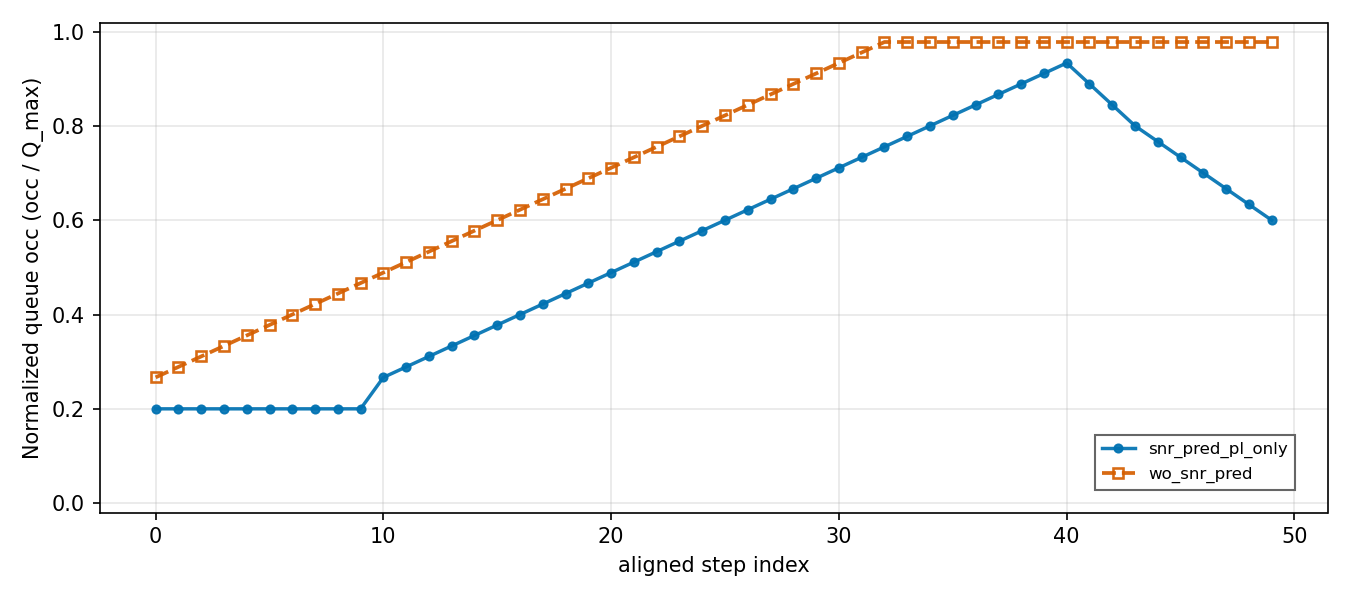}
    \label{fig:ab_occb_qnorm}
  }
  \hfill
  \subfloat[Qualified Frames per step]{
    \includegraphics[width=0.85\linewidth]{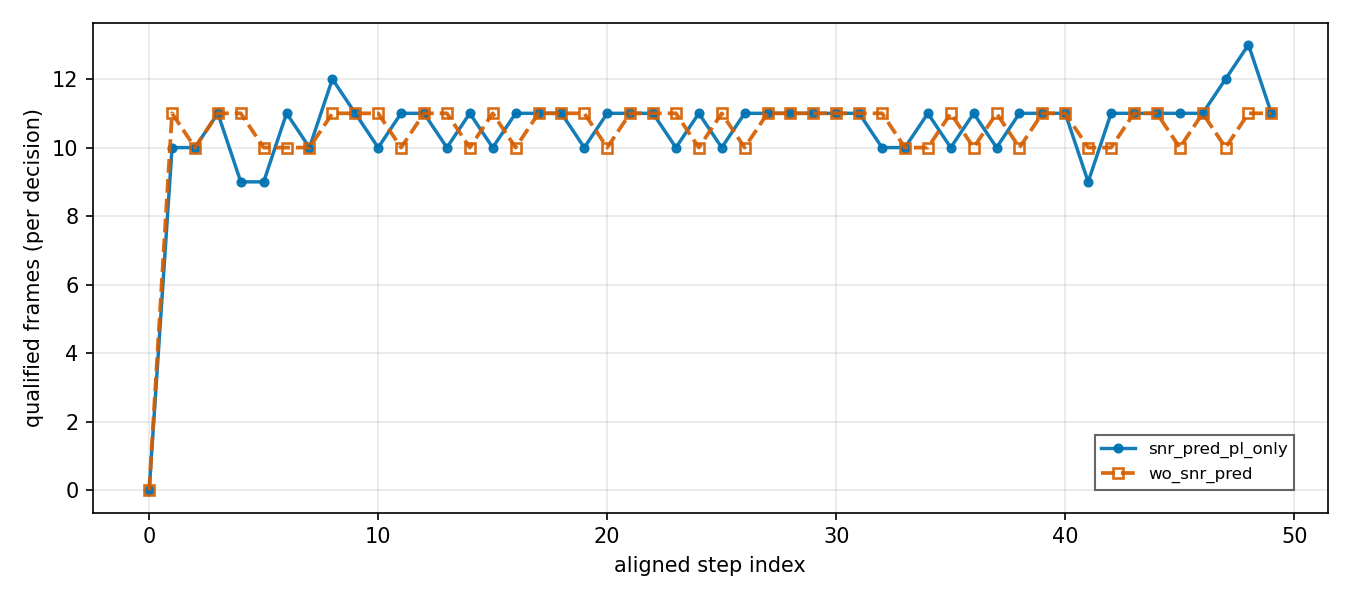}
    \label{fig:ab_qual_frames}
  }
  \caption{Impact of SNR predict on performance of proposed RL policy.}
  \label{fig:ablation}
\end{figure}

To investigate the practical impact of SNR prediction, we train two policies — one with the SNR predictor module (\texttt{snr\_pred\_pl\_only}) and one without (\texttt{wo\_snr\_pred}) — and compare their performance in terms of target rate adaptation and the final number of qualified frames. As shown in Fig.~\ref{fig:ablation}, the policy without SNR prediction selects the $C = 128$ level and keeps it unchanged, implying that no rate switching is performed over the entire SNR sweep. As a result, the transmission buffer becomes full at step $32$, leading to a small number of dropped frames ($17$ frames). In contrast, the policy equipped with SNR prediction proactively adjusts the target rate based on the forecasted SNR, thereby avoiding buffer overflow and achieving a higher number of qualified frames. This confirms that SNR prediction enables timely rate adaptation, prevents congestion, and improves overall transmission efficiency.

\subsubsection{Ablation Study: Effect of SNR Prediction on Channel ModNet}
We compare two RL-based policies: one that feeds the predicted SNR to both the RL agent and the Channel ModNet, denoted as \texttt{snr\_pred\_encoder}, and the other that supplies the predicted SNR only to the RL agent (while the Channel ModNet uses the instantaneous SNR), denoted as \texttt{snr\_pred\_pl\_only}. This isolates the impact of predictive channel information on the encoding adaptation process. As shown in Table~\ref{tab:ablation_snr_pred}, no significant performance difference is observed between the two variants under the current smoothly varying SNR scenario. The primary reason is that the SwinJSCC encoder and the Channel ModNet have already been fine-tuned to accommodate the entire SNR range of interest. In our experiments, the SNR trajectory generated by the link budget model changes very gradually over time, and the prediction horizon does not provide sufficient additional gain to meaningfully affect the reconstruction quality. Consequently, the benefit of feeding predicted SNR into the Channel ModNet is marginal in this setting, while its contribution to the RL agent’s rate decision remains essential for avoiding buffer overflow and improving throughput.

\begin{table*}[!t]
\caption{Ablation study on the impact of SNR prediction.\label{tab:ablation_snr_pred}}
\centering
    \begin{tabular}{|c||c|c|c|}
        \hline
        \textbf{Metrics} & \texttt{wo\_snr\_pred} & \texttt{snr\_pred\_pl\_only} & \texttt{snr\_pred\_encoder}\\
        \hline
        qualified & 521 & 524 & 524\\
        \hline
        forwarded & 539 & 564 & 564\\
        \hline
        dropped & 17 & 0 & 0\\
        \hline
        mean\_channel\_number & 128.00 & 115.84 & 115.84\\
        \hline
        cbr\_at\_fwd & 0.0833 & 0.0762 & 0.0762\\
        \hline
        qual/fwd (\%) & 96.660 & 92.908 & 92.908\\
        \hline
    \end{tabular}
\end{table*}

\section{Conclusions}
\label{sec:conclusion}
In this paper, we have proposed an adaptive semantic communication framework for LEO satellite-to-ground image transmission based on SwinJSCC. To address the dynamic nature of the satellite link and limited on-board resources, we introduced a transmission buffer and a link budget model to simulate SNR evolution over an overpass. A reinforcement learning agent dynamically selects the compression ratio at each decision interval, supported by a lightweight polynomial-based SNR predictor that compensates for control loop latency.

Simulation results demonstrate that the proposed RL-based policy significantly outperforms fixed-rate baselines in terms of qualified frame rate and buffer management, achieving no packet loss. SNR prediction proves vital for proactive rate adaptation, preventing overflow and improving throughput. The framework effectively balances transmission efficiency and reconstruction quality, offering a promising solution for real-time multi-spectral image downlink on resource-constrained LEO satellites.

Future work includes extending the framework to multi-overpass scenarios, incorporating realistic channel impairments such as rain fading and scintillation, and exploring more advanced RL algorithms.

\bibliographystyle{IEEEtran}
\bibliography{ref_routing,ref_semc}

\vfill

\end{document}